\documentclass[reprint,amsmath,amssymb,aps,prl,superscriptaddress,showkeys]{revtex4-2}
\usepackage{graphicx}
\usepackage{dcolumn}
\usepackage{bm}
\usepackage[svgnames,x11names,table]{xcolor}
\usepackage{hyperref}
\hypersetup{
	colorlinks=true,
	linkcolor=Blue4,
	filecolor=Blue4,
	citecolor=Blue4,      
	urlcolor=Blue4,
}

\usepackage{ragged2e}

\usepackage{xcolor}
\usepackage{ulem}
\usepackage{orcidlink}

\usepackage{caption}    
\usepackage{subcaption} 
\begin{document}

\title{Coherent Two-State Oscillations in False Vacuum Decay Regimes}

\author{Peiyun Ge}\thanks{These authors contributed equally to this work.}
\affiliation{State Key Laboratory of Low Dimensional Quantum Physics, Department of Physics, Tsinghua University, Beijing 100084, China}

\author{Xiao Wang\,\orcidlink{0000-0002-3022-7260}}
\thanks{These authors contributed equally to this work.}
\email{xw970921@gmail.com}
\affiliation{State Key Laboratory of Low Dimensional Quantum Physics, Department of Physics, Tsinghua University, Beijing 100084, China}
\affiliation{Beijing Academy of Quantum Information Sciences, Beijing 100193, China}

\author{Yu-Xin Chao}
\affiliation{State Key Laboratory of Low Dimensional Quantum Physics, Department of Physics, Tsinghua University, Beijing 100084, China}

\author{Rong Lu}
\affiliation{State Key Laboratory of Low Dimensional Quantum Physics, Department of Physics, Tsinghua University, Beijing 100084, China}

\author{Li You}
\affiliation{State Key Laboratory of Low Dimensional Quantum Physics, Department of Physics, Tsinghua University, Beijing 100084, China}
\affiliation{Beijing Academy of Quantum Information Sciences, Beijing 100193, China}
\affiliation{Frontier Science Center for Quantum Information, Beijing 100084, China}
\affiliation{Hefei National Laboratory, Hefei, Anhui 230088, China}

\date{\today}

\begin{abstract}
Coherent two-state oscillations are observed in numerical simulations of the one-dimensional transverse-longitudinal-field Ising model (TLFIM) within false vacuum decay regimes. Starting from the false vacuum (a nearly fully polarized ferromagnetic state), we show that in moderate-sized systems, at resonances $h\approx 2J/n$ (with longitudinal field $h$, transverse field $J$, and an integer $n$), the expected decay can give way to coherent oscillations between the false vacuum and a symmetric resonant state.
The oscillation frequency, i.e., the tunneling splitting, is observed notably to exhibit a superradiant-like $\sqrt{L}$ enhancement, as confirmed by a Schrieffer-Wolff analysis. In large chains, coherence remains for $n\gtrsim L/2$ due to bubble-size blockade and is robust against stronger transverse fields; for small $n$, long-range interactions can stabilize the oscillations by lifting multi-bubble degeneracies, establishing a robust many-body coherence mechanism beyond perturbative and finite-size limits.

\end{abstract}

\maketitle

\textit{Introduction.} 
False vacuum decay (FVD) \cite{PhysRevD.15.2929,PhysRevD.16.1762,devoto2022false}, the decay of a metastable state via nucleation and growth of true-vacuum bubbles, admits a microscopic mechanism well captured by spin models in terms of domain walls \cite{Bray1994} due to short-range interactions.
 This mechanism has far-reaching implications in cosmology \cite{PhysRevD.21.3305,Guth1981,Vilenkin1984}, particle physics \cite{Isidori2001,Buttazzo2013}, and non-equilibrium dynamics \cite{PhysRevLett.21.973,Fialko2015,Braden2019,Billam2019,Batini2024}. While challenging to study in high-energy experiments, this phenomenon has recently become possible to emulate in NISQ quantum simulators across a variety of platforms \cite{zhu2024,Zenesini2024False,luo2025,vodeb2025stirring}. 

 \begin{figure*}[th!]
  \centering
  \includegraphics[width=\textwidth]{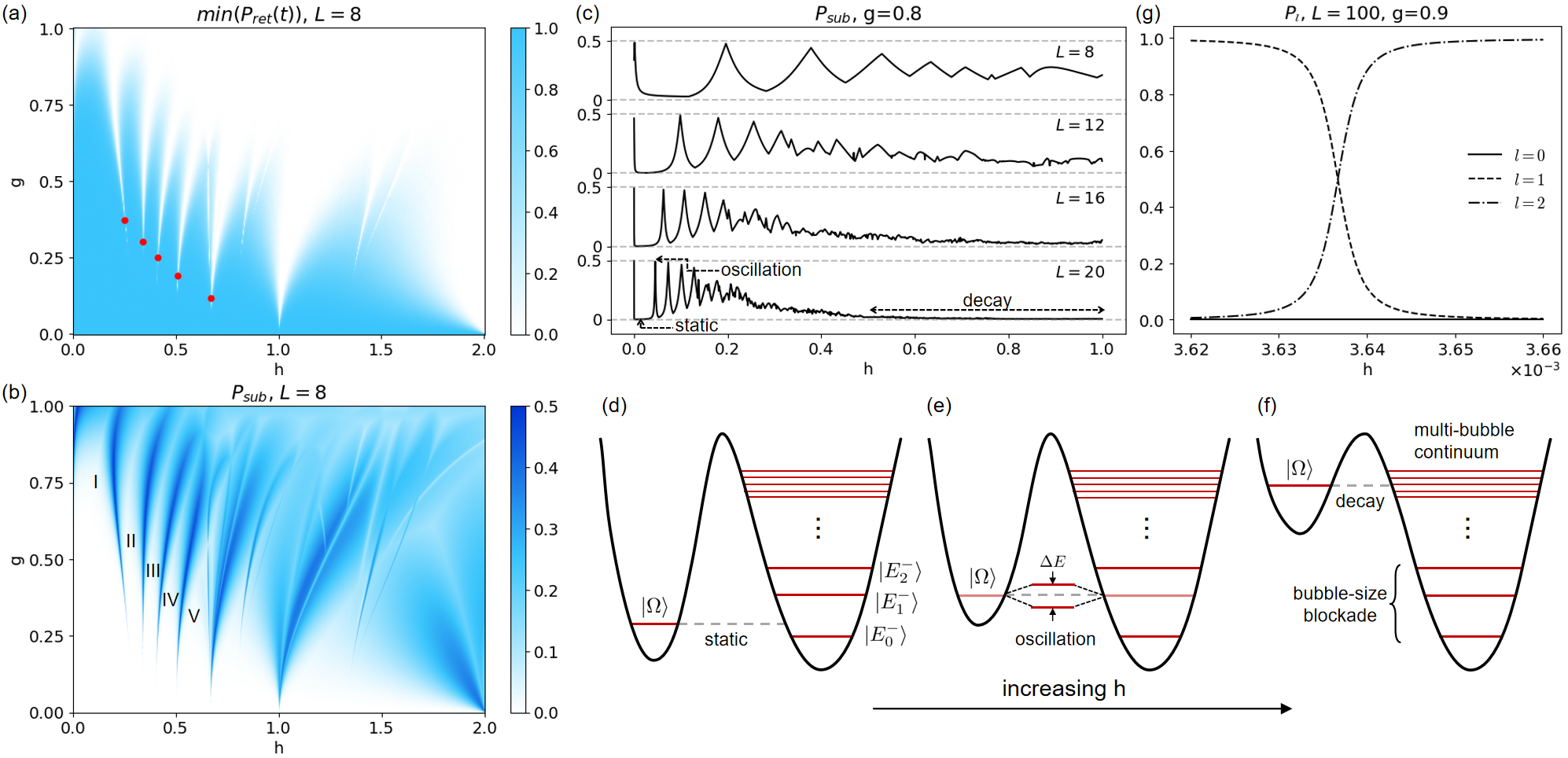} 

  \caption{(a) The minimal return probability $\min \left(P_{\text{ret}}(t) \right)$ from the false vacuum state $\vert \Omega \rangle$ for $L=8$, shown as a function of the Hamiltonian parameters $h$ (x-axis) and $g$ (y-axis). (b) The same as in (a), but for the sub-leading overlap $P_{\text{sub}}$ between the false vacuum state $\vert \Omega \rangle$ and the eigenstates of the Hamiltonian (\ref{eq.H}). The blue regions mark the parameters supporting coherent 2-level dynamics. (c) $P_{\text{sub}}$ at fixed $g=0.8J$ for $L=8,12,16,$ and $20$. \textcolor{black}{(d)--(f) Schematic evolution of the spectrum with increasing $h$, revealing the three dynamical regimes highlighted in (c). (g) Overlaps $P_l$ between the false vacuum state and the low-lying eigenstates $|E_l\rangle$ ($l=0,1,2$) for $L=100$ with $g=0.9J$.}}
  \label{fig:phasediagram}
\end{figure*}

The transverse-longitudinal-field Ising model (TLFIM) provides a paradigm for studying false vacuum decay. Following Ref. \cite{Lagnese2024}, we define the false vacuum state $\vert \Omega\rangle$ as the symmetry-breaking ferromagnetic ground state of TLFIM at vanishingly small longitudinal field $h$. For weak transverse field $g$, the false vacuum is well approximated by the ferromagnetic all-spin-up state $\vert \Omega\rangle \! \approx \! \vert \! \uparrow,\uparrow,...,\uparrow\rangle$. Earlier studies \cite{PhysRevB.104.L201106,Zenesini2024False,PhysRevB.110.155103,PhysRevB.60.14525,Lencses2022} reveal that false vacuum decay proceeds resonantly in the TLFIM when the energetic cost of nucleating a bubble of flipped spins vanishes. Specifically, creating a bubble of size $n$ requires forming two domain walls and costs the Ising energy $4J$, while simultaneously flipping the $n$ spins of a bubble gains a Zeeman energy $2nh$ from the longitudinal field $h$. Resonant tunneling occurs when $2J/h$ approaches an integer $n$, where $n$ defines the \textit{resonant bubble size} (RBS) order. The condition $2J/h=n$ not only enables resonant coupling between the false vacuum $\vert \Omega\rangle$ and single-bubble states, it also facilitates sequential resonant tunneling between single-bubble and multi-bubble states \cite{vodeb2025stirring}, i.e., states containing more than one resonant-size bubble. Such sequential tunneling leads to an exponential decay of the Loschmidt echo \cite{PhysRevLett.96.140604,Sharma_2014, Mendoza-Arenas_2022}, i.e., the return probability of the initial false vacuum state behaves like \(P_{\text{ret}}(t)\propto e^{-\Gamma t}\). This interesting decay behavior is captured by both the Langer’s droplet expansion theory \cite{Langer1969,PhysRevB.60.14525,Richards1996} and the Callan-Coleman multi-bounce formalism \cite{devoto2022false,Carosi2025,PhysRevD.16.1762}, or by the more recent Gaussian ansatz approaches \cite{johansen2025,maertens2025}. In the thermodynamic limit, the  resonance with multi-bubble states mentioned above is understood to trigger a many-body cascade that drives the incoherent decay of the false vacuum.

\textcolor{black}{In single-particle physics, resonant tunneling between two nearly degenerate states produces coherent oscillations with frequency set by the tunneling splitting $\Delta E$ \cite{RevModPhys.59.1}. In FVD, on the other hand, the resonant false vacuum level is typically embedded in a dense multi-bubble manifold; hybridization with this manifold rapidly dephases the dynamics and renders tunneling irreversible. 
Considerable efforts have been devoted to observing coherent oscillations or persistent recurrences in interacting many-body systems \cite{Bernien_2017,Bluvstein_2021,Zhang_2017,Choi_2017,Wu_2024}.
A promising route suggests employing long-range interactions \cite{landig-2016} to restructure the spectrum, thereby stabilizing non-equilibrium coherence that is absent otherwise  in short-range interaction settings \cite{PhysRevX.15.011020}. Hence, the landscape of FVD may admit regimes beyond simple irreversible decay.}

Here in this Letter, we report a striking numerical observation that deviates from the conventional false vacuum decay scenario in the TLFIM. Specifically, for small systems at weak transverse fields, coherent two-state oscillations between the false vacuum $|\Omega\rangle$ and a symmetric state $|S_n\rangle$ are detected, manifested by periodic revivals of the return probability and a sub-leading eigenstate overlap $P_{\text{sub}} \sim 0.5$. Furthermore, the effective oscillation frequency or tunneling splitting is found to exhibit a superradiance-like $\sqrt{L}$ enhancement, captured by constructing a Schrieffer-Wolff transformation (SWT) in the symmetric subspace. For larger systems, we identify two mechanisms that can sustain coherent oscillations: (i) the bubble size blockade effect at $ n\gtrsim L/2$, which suppresses multi-bubble states via spatial constraints and promotes robustness against interaction disorder, and (ii) long-range Ising interactions that lift the degeneracy between single- and multi-bubble manifolds. 
Both sustain robust many-body coherence beyond perturbative and finite-size limits, and suggest practical strategies for quantum simulation and engineering of recurrences in metastable dynamics.

\textit{Model.}
The TLFIM on a ring of length $L$ with periodic boundary conditions is given by
\begin{equation}\label{eq.H}
    \hat{H} = -J \sum_{i=1}^{L} \hat{\sigma}_i^z \hat{\sigma}_{i+1}^z - g \sum_{i=1}^{L} \hat{\sigma}_i^x + h \sum_{i=1}^{L} \hat{\sigma}_i^z ,
\end{equation}
where $\hat{\sigma}_i^z$ and $\hat{\sigma}_i^x$ are Pauli matrices acting on site $i$, the periodic boundary condition assures $\hat{\sigma}_{L+1}^z \equiv \hat{\sigma}_1^z$, and $J$ is the strength of the nearest-neighbor (NN) Ising interaction, $g$ is the transverse field strength, $h$ is the longitudinal field strength. Throughout this work, we set $J=1$ as the unit of energy and focus on the ferromagnetic regime $g<J$.

Starting from the false vacuum state $\vert \Omega\rangle$, obtained as the ground state of the TLFIM with a tiny longitudinal field (e.g., $h=-0.01$) via exact diagonalization, the dynamics under Hamiltonian (\ref{eq.H}) is tracked at a positive $h$ \footnote{Although the definitions of the false vacuum in Refs.~\cite{PhysRevB.104.L201106} and \cite{Lagnese2024} differ slightly, they yield essentially the same results in our work.}. 

First, in Fig~\ref{fig:phasediagram}(a), we present the minimum return probability, $\min \left(P_{\text{ret}}(t) \right)$, where $P_{\text{ret}}(t)\equiv \vert \langle \Omega \vert e^{-i \hat{H} t} \vert\Omega\rangle \vert^2$, over a time window of $400$ (in units of $J^{-1}$). Figure~\ref{fig:phasediagram}(a) clearly distinguishes whether the system lies within the false vacuum decay regime: in this regime, FVD results in a vanishingly small return probability. The observed regimes occur around $h=2J/n$ for various RBS order $n\le L-1$.
 Next, in Fig.~\ref{fig:phasediagram}(b), we show the sub-leading overlap $P_{\text{sub}}$, defined as the second-largest value among the overlaps $P_{l}=\vert \langle \Omega \vert E_l\rangle \vert^2$ between the initial false vacuum state $|\Omega\rangle$ and the eigenstates $|E_l\rangle$ of the Hamiltonian in Eq.~(\ref{eq.H}).
In the vicinity of $P_{\text{sub}}\!\!=\!0.5$, the system’s evolution is governed predominantly by just two eigenstates, corresponding to persistent coherent oscillations between these two many-body states. 
Specifically, for $L=8$, coherent oscillations appear only when $n\geq 3$, with oscillation period $T$ scaling exponentially with respect to $n$, or $T\sim 2\pi/g^{n}$. The red dots in Fig.~\ref{fig:phasediagram}(a) mark the corresponding parameters for high-fidelity two-state oscillations ($P_{\text{sub}}>0.49$) at a period around $T\approx600$.

In Fig.~\ref{fig:phasediagram}(c), we fix the transverse field at a representative value $g=0.8$ and plot $P_{\text{sub}}$ versus $h$ for increasing $L$. For a finite system size $L$, as the longitudinal field $h$ increases from $0$, we find the false vacuum decay dynamics alternates between single-state stationary evolution and two-state oscillations, respectively associated with $P_{\text{sub}}\!\sim\!0$ and $P_{\text{sub}}\!\sim\!0.5$. For $L=20$, this cycle of change occurs about four times before the system enters the false vacuum decay regime.

\textcolor{black}{In Fig.~\ref{fig:phasediagram}(d)--(f), we show schematically the spectrum for different dynamical phases as the symmetry-breaking field $h$ is increased. For low-lying eigenstates, the spectrum exhibits a $\mathbb{Z}_2$-symmetry-broken structure. The eigenstates separate into two sectors according to the sign of their energy shift with increasing $h$. We denote by $\{\vert E^-_m\rangle\}$ the translationally invariant eigenstates whose energy decreases as $h$ increases, as illustrated in the right sector in Fig.~\ref{fig:phasediagram}(d)--(f). In the complementary sector where the energy increases with $h$, the lowest-energy eigenstate is denoted as $\vert E^+_0\rangle$. In the static phase (d), $\vert E^+_0\rangle$ is sufficiently detuned from all $\vert E^-_m\rangle$ and is well approximated by the false vacuum $\vert\Omega\rangle$ as a result of low susceptibility explained in Sec.~I of the supplemental material (SM) \cite{SM}. Consequently, the quenching dynamics initiated from $\vert\Omega\rangle$ remain essentially frozen.
Upon increasing $h$, $\vert\Omega\rangle$ is tuned into resonance with a particular $\vert E^-_m\rangle$, and the two hybridize into a pair of eigenstates, giving rise to coherent oscillations that define the phase (e). 
This effective two-state oscillation occurs only for low-lying $\vert E^-_m\rangle$, where a blockade mechanism, to be analyzed later in this work, ensures the absence of nearby levels; at higher energies the spectrum becomes dense and forms a multi-bubble continuum once the blockade is lifted. Upon further increasing $h$, the system enters the decay phase (f), where the level associated with $\vert \Omega\rangle$ is pushed into the multi-bubble continuum and hybridizes with a massive number of eigenstates, so false vacuum decay overtakes coherent oscillations.} 

\textcolor{black}{Consistent with the above picture, the oscillation regions with $P_{\text{sub}}\approx 0.5$ in Fig.~\ref{fig:phasediagram}(b) form narrow strips within the non-static regime of Fig.~\ref{fig:phasediagram}(a), separating the static regions I, II, III, \dots. Within each strip, the dynamics is governed by the tunneling splitting $\Delta E$ between the hybridized eigenstates $\frac{1}{\sqrt{2}}(\vert\Omega\rangle\pm\vert E^-_m\rangle)$.} 
The coherent two-state evolution can be approximately described by 
\begin{equation}\label{eq.two-state-bloch-oscillation}
\begin{split}
    e^{-i\hat{H}t} \vert \Omega\rangle &\approx e^{-i \nu t} \bigg[ \cos \left(\frac{\Delta E}{2} t\right) \vert \Omega\rangle  - i \sin\left(\frac{\Delta E}{2} t\right) \vert  E^-_m \rangle \bigg],
\end{split}
\end{equation}
with overall phase $\nu$ and \textcolor{black}{oscillation period $T=2\pi/\Delta E$.}

\textcolor{black}{The picture of tunneling splitting for the two-state oscillation holds beyond the limit of weak transverse field and the small chain. Figure~\ref{fig:phasediagram}(g) presents the numerical results for $L=100$ at $g=0.9J$.
The overlaps $P_l=\lvert\langle\Omega \vert E_l\rangle\rvert^2$ between $\vert \Omega\rangle$ and the low-lying eigenstates of the post-quench Hamiltonian $\hat{H}$ are
computed using MPS-based DMRG \cite{White1992,White1993,tenpy}. As the symmetry-breaking field $h$ is increased, the dominant weight of the initial state $\vert \Omega \rangle$ transfers from the first to the second excited eigenstate of the spectrum, signaling an avoided level crossing between $\vert \Omega\rangle$ and $\vert E^-_1\rangle$, as depicted in Fig.~\ref{fig:phasediagram}(e). Around crossing, the sub-leading overlap $P_{\text{sub}}$ of the initial state reaches $0.5$, indicating two-state oscillations even in this moderate-sized chain at a strong transverse field. }

\textit{Analytic expressions for resonant states at small $g$.}
The vertical blue strips in Fig.~\ref{fig:phasediagram}(b) indicate that the nontrivial resonant states $\vert E^-_m\rangle$ at $g=0.8$ [corresponding to the peaks of $P_{\text{sub}}$ in Fig.~\ref{fig:phasediagram}(c)] continuously connect, as $g\!\to\!0$, to analytically tractable states. In this limit $\vert E^-_m\rangle$ reduces to the symmetric single-bubble states $\vert S_n\rangle$ \textcolor{black}{with $n=L-m$}. Meanwhile, in this weak-$g$ regime the false vacuum $\vert \Omega\rangle$ is well approximated by $\vert\uparrow,\uparrow,\ldots,\uparrow\rangle$.

\begin{figure}[t!]
  \centering
  \includegraphics[width=0.48\textwidth]{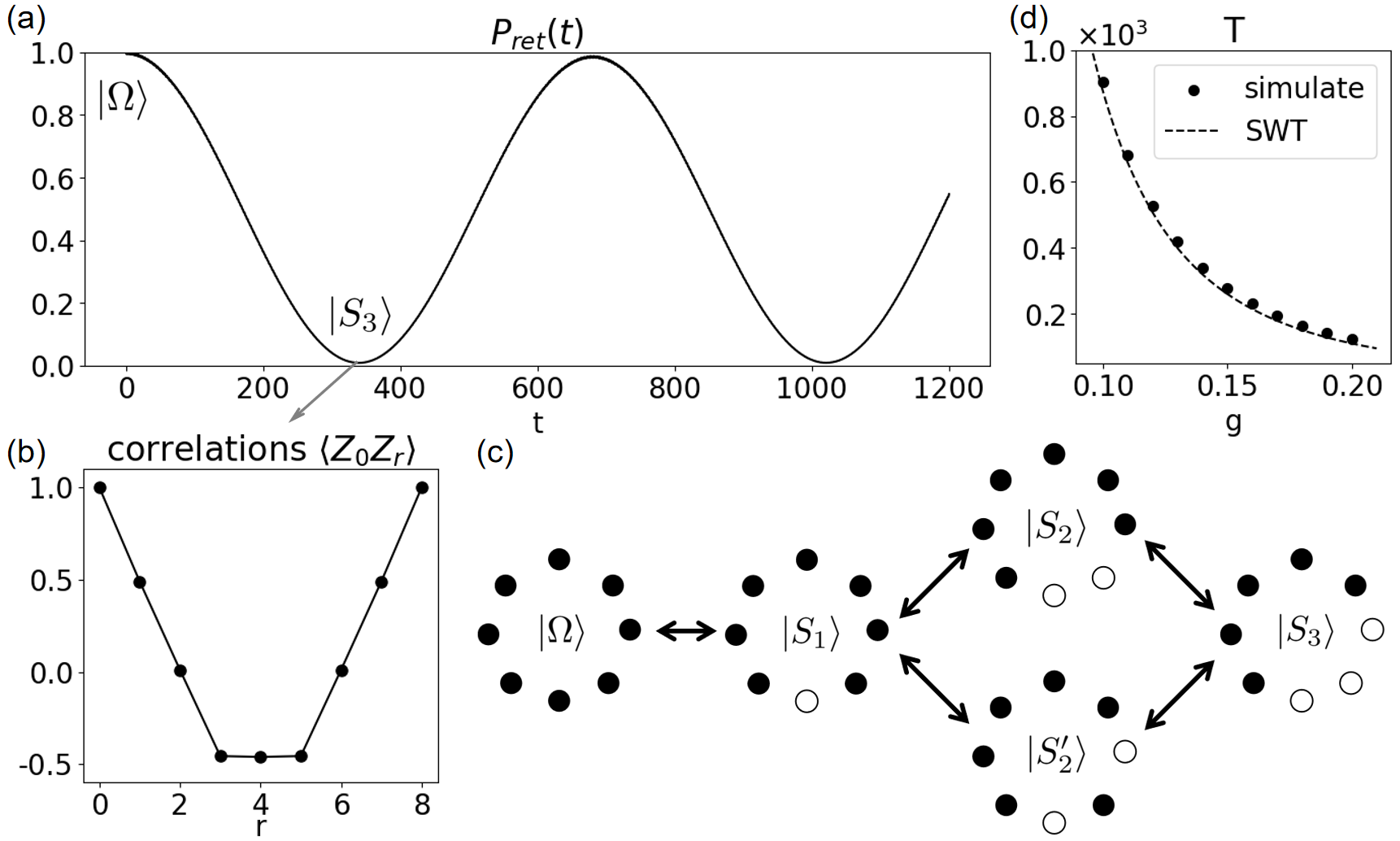} 

  \caption{The evolution starting from the false vacuum state $\vert\Omega\rangle$ at $L=8$, $h=0.67$, and $g=0.11$. (a) The return probability to the initial state. 
  (b) The ZZ correlator. 
  (c) The five translationally invariant states closely support the oscillation. \textcolor{black}{(d) The simulated and the SWT-predicted oscillation period $T$ for various $g$.}
  }
  \label{fig:all}
\end{figure}

In Fig.~\ref{fig:all}, the evolution of the initial false vacuum state $\vert\Omega\rangle$ at $g=0.11$ and $h=0.67$ is shown. It features a coherent two-state oscillation inside the false vacuum decay regime, with RBS order $n=2J/h=3$. The return probability $P_{\text{ret}}(t)$ oscillates between 1 and 0 with a period around $680$ in Fig.~\ref{fig:all}(a).
In Fig.~\ref{fig:all}(b), the two-site correlations $\langle \hat{Z}_0 \hat{Z}_r \rangle$ show that the resonant state is ferromagnetic at short range but anti-ferromagnetic at long range. 

The above numerical results suggest that, at a small transverse field $g=0.11$, the state $\vert  E^-_m \rangle$ resonating with the false vacuum $\vert \Omega \rangle$ in Fig.~\ref{fig:all} is well approximated by
\begin{equation}\label{eq.S-state}
    \vert S_3\rangle = \frac{1}{\sqrt{L}} \sum\limits_{i} \hat{S}^-_{i-1} \hat{S}^-_{i} \hat{S}^-_{i+1} \vert\Omega\rangle,
\end{equation}
where $\hat{S}^-_{i}\equiv \vert\! \downarrow \rangle_i \langle \uparrow\! \vert$ flips the up spin down at site $i$. The symmetric state $\vert S_3\rangle$ contains a single bubble of length 3 and remains invariant under site translation $i\to i+1$ on the $L$-site ring. The correlations in Fig.~\ref{fig:all}(b) can be completely explained by Eq.~(\ref{eq.S-state}) \footnote{The bubble-size Bloch oscillation \cite{10.21468/SciPostPhys.12.2.061} is absent in the coherent two-state oscillation described by Eq.~(\ref{eq.two-state-bloch-oscillation}). Such Bloch oscillations would appear when the system is initialized in the all-up state $\vert\!\uparrow,\uparrow,...,\uparrow\rangle$, rather than in the false vacuum state $\vert\Omega\rangle$.}, as detailed in SM \cite{SM} Sec.~II. Next, leveraging the analytical properties in the small $g$ limit, we uncover a collective enhancement of the oscillation frequency (or equivalently, the tunneling splitting) $\Delta E$.

\textit{The enhanced oscillation frequency.}
At the false vacuum decay resonance $h=2J/n$ of small $g$, an effective two-state model was developed previously based on Schrieffer-Wolff transformation (SWT) \cite{PhysRevB.103.L220302}. For $n=3$ or $h=2J/3$, it gave \cite{PhysRevB.103.L220302} 
\begin{equation}\label{eq.two-state-approx-n=3}
H_{\mathrm{eff}} \approx E_0\left(h\right)+\left[\begin{array}{cc}
0 & -\frac{81}{64} g^3 \\
-\frac{81}{64} g^3 & 4-6 h
\end{array}\right] .
\end{equation}
However, this result does not quantitatively reproduce the observed oscillatory frequency $\Delta E$ in Eq. (\ref{eq.two-state-bloch-oscillation}). Moreover, Ref.~\cite{vodeb2025stirring} produces a complicated projected many-body Hamiltonian given by the leading order SWT [see Eq.(25) in its SM]. Since the all-up state is an eigenstate of the corresponding many-body Hamiltonian, it cannot account for the two-state coherent oscillations observed here.

We instead perform SWT to the third order in $g$ within the subspace spanned by the five symmetric states (as illustrated in Fig.~\ref{fig:all}).
The resulting effective two-state Hamiltonian consequently reads (see SM \cite{SM} Sec.~III)
\begin{equation}\label{eq.sym-swt-n=3}
H_{\mathrm{eff}}^{mod} \approx E_0\left(h,g\right) + \left[\begin{array}{cc}
\Delta(h,g,L) & -\kappa(h)\sqrt{L}g^3 \\
-\kappa(h)\sqrt{L}g^3 & 4-6 h 
\end{array}\right] ,
\end{equation}
where $\Delta$ is an $O(g^2)$ self-energy shift of the false vacuum state $\vert \Omega\rangle$, and $\kappa(h)=\frac{81}{64}$ at $h=\frac{2}{3}$. Around the $n=3$ resonance,
$\vert \Omega\rangle$ strongly hybridizes with $\vert S_3\rangle$.
The effective Rabi frequency \textcolor{black}{[proportional to the tunneling splitting $\Delta E$ in Eq.~(\ref{eq.two-state-bloch-oscillation})]} scales as $\sqrt{L}$, characteristic of a superradiance-like collective enhancement. A similar $\sqrt{L}$ scaling was earlier experimentally reported in fully Rydberg-blockaded systems \cite{Labuhn:2016xba}, when only one atom can be excited \footnote{apart from the fully blockade case, Ref. \cite{Labuhn:2016xba} also explored the nearest neighbor blockade region in their Fig. 3b, corresponding to the $n=1$ RBS order. However, in Fig.~\ref{fig:phasediagram}(b), we see that $n=1$ cannot sustain two-state coherence.}. Our modified effective Hamiltonian (\ref{eq.sym-swt-n=3}) is more general, and gives the oscillation period $\frac{\pi}{\kappa \sqrt{L}g^3}\sim 659$ at $h=\frac{2}{3}$, which corroborates well with the numerical simulations. \textcolor{black}{In Fig.~\ref{fig:all}(d), we show that the oscillation period predicted by Eq.~(\ref{eq.sym-swt-n=3}) agrees well with simulations as $g$ increases along the $n=3$ strip in Fig.~\ref{fig:phasediagram}(b), until the breakdown of two-state oscillations at $g\gtrsim0.2J$.}

\begin{figure}[th!]
    \centering
    \includegraphics[width=1.0\linewidth]{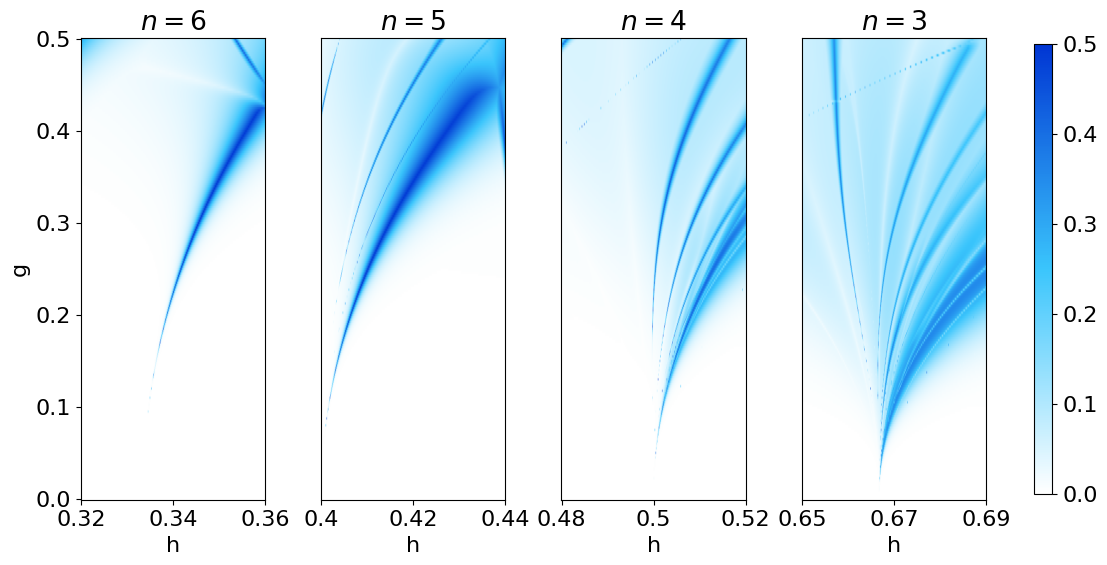}
    \caption{The sub-leading overlap between the false vacuum state and the eigenstates of the Hamiltonian (\ref{eq.H}) for $L=12$ around different RBS orders $n$. }
    \label{figL12n3456}
\end{figure}

\textit{The bubble size blockade effect.}
Figure~\ref{fig:phasediagram}(c) shows that the coherent two-state oscillation persists at large RBS orders $n\gtrsim L/2$. We interpret this behavior as the bubble size blockade effect, \textcolor{black}{highlighted in Fig.~\ref{fig:phasediagram}(f)} and discussed below. Around the RBS order $n$ (at $h\approx2J/n$) 
, a bubble with size $n$ resonates with the false vacuum, but the states with multiple bubbles are also at resonance, \textcolor{black}{resulting in the multi-bubble continuum as sketched in Fig.~\ref{fig:phasediagram}(f).} The participation of this continuum in the dynamics leads to the thermalization in FVD. However, as $n$ approaches $L$, the multi-bubble states can no longer fit into the ring. Thus the dynamics is restricted within the subspace of the false vacuum and the one-bubble state \textcolor{black}{ $\vert E^-_m\rangle$, as shown in Fig.~\ref{fig:phasediagram}(e)}, hence a coherent two-state oscillation overtakes the decay behavior. 
This effect is examined in Fig.~\ref{figL12n3456}. In a ring with $L=12$, we plot the $P_{\text{sub}}$ around RBS orders $n=3\sim6$. At larger $n=5,6$, near-perfect coherence is observed with $P_{\text{sub}}$ approaching 0.5, according to the bubble size blockade. Around smaller orders $n=3,4$, the blockade effect vanishes because states with 2 or 3 bubbles can participate in the dynamics. Consequently, as $L$ increases from $8$ to $12$, while the overlap $P_{\text{sub}}$ still exhibits an oscillating regime (in blue) around $n=3$, it no longer approaches 0.5. 

\begin{figure*}[th!]
    \centering
  \includegraphics[width=1\textwidth]{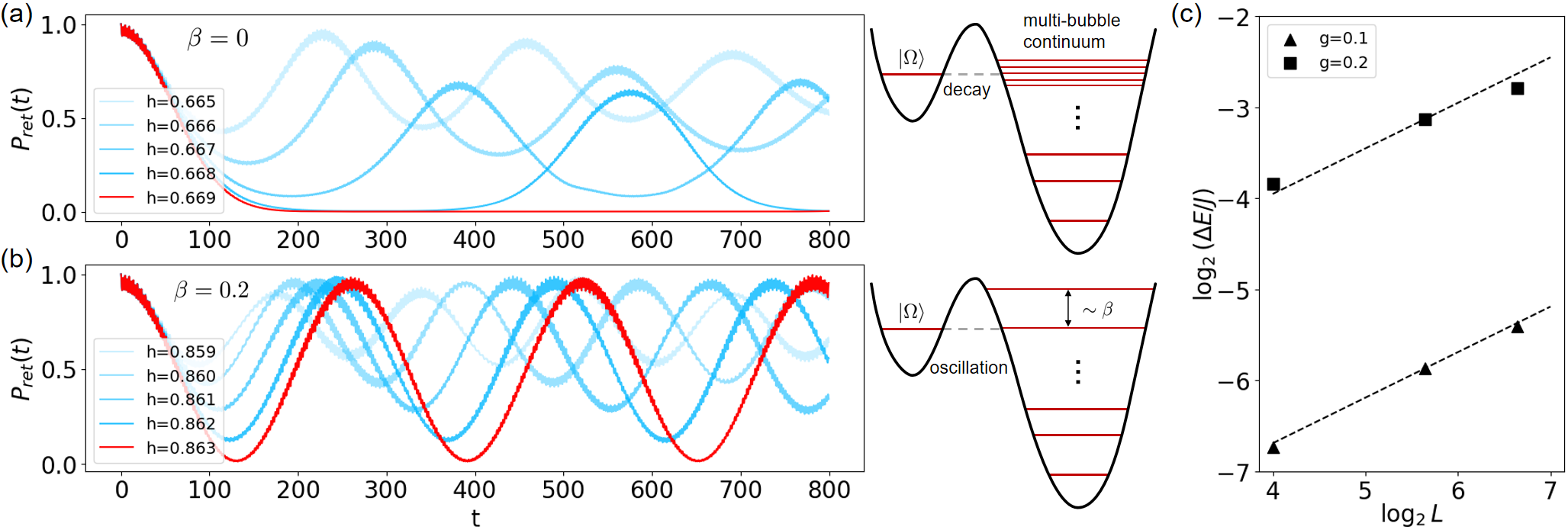} 
  \caption{\textcolor{black}{Dynamics with global-range interaction of strength $\beta$. (a) The dynamics of an $L=100$ ring around RBS order $n=3$. The Loschmidt echo of the state at $g=0.1$, $\beta=0$, as $h$ approaches the resonant value for $n=3$. The red curve denotes the resonant $h$, while the blue curves represent the different nearby $h$ approaching this resonance. The system undergoes decay dynamics at resonance. (b) Same as (a), but for $\beta=0.2$. As $h$ approaches resonance, the system stays in two-state oscillation. (c) Scaling of the tunneling splitting $\Delta E=2\pi/T$ with respect to $g$ and $L$ at RBS order $n=3$, extracted from real-time evolution. Data are shown for $g=0.1,0.2$ and $L=16,50,100$ at $\beta=0.2$. Dashed lines are guides to the eye with the expected superradiant slope $1/2$, anchored at the $L=50$ data point.}}
  \label{fig:globalrange}
\end{figure*}

At even larger $L$, when $n\gtrsim L/2$, we expect the same blockade effect (provided that the transverse field $g$ is not too large to break it), which serves as the underlying reason for the alternations of $P_{\text{sub}}$ shown in Fig.~\ref{fig:phasediagram}(c). As $n$ increases at fixed $g$, the resonant (blue) regime in the diagram of $P_{\text{sub}}$ becomes exponentially narrower, making it increasingly difficult to identify parameters supporting two-state oscillations. Nevertheless, as $g$ approaches $J$, the problem of exponential narrowing is mitigated, \textcolor{black}{as illustrated in the SM \cite{SM} Sec.~IVA}. Figure~\ref{fig:phasediagram}(c) and (g) further shows that the bubble size blockade effect persists even at such large $g$, where the all-up state no longer approximates the false vacuum. In particular, for a moderate chain size $L = 20$, coherent two-state oscillations can still be observed at $n=L-1$ and $g=0.8$, and the same is expected for $L=100$. Moreover, for the blockade-facilitated two-state oscillations in this large-$g$ regime, we find these oscillations are robust against interaction disorder, as shown in the SM \cite{SM} Sec.~IVB. This robustness is suspected to originate from the non-local nature of domain walls
at large $g$
\footnote{A numerical study of this non-local behavior of domain walls is given in Ref.~\cite{MilstedPRX}.}, which effectively averages out local fluctuations in the interaction strength.

\textit{Lifting the degeneracy of the multi-bubble manifold.} At large $L$, two-state dynamics can again arise, not necessarily at a large $n$, but by introducing long-range Ising interactions, with which the interaction between bubbles can no longer be neglected. As a result, the state degeneracy between the single-bubble manifold and the multi-bubble manifold is lifted, as sketched in Fig.~\ref{fig:globalrange}. The multi-bubble states are thus decoupled from the oscillatory dynamics between $\vert \Omega\rangle$ and $\vert E_m^-\rangle$, even outside the bubble-size blockade regime. 
\textcolor{black}{In the SM Sec.~V, we show that the coherent oscillations at RBS order $n=3$ are stabilized by an $r^{-3}$ Ising interaction in an $L=12$ ring.}

 For even larger systems, we can introduce global-range interactions to stabilize coherent two-state oscillations. As detailed in SM \cite{SM} Sec.~VI, this can be realized by adding a cavity-mediated spin-squeezing term $- \frac{\beta J}{2L} ( \sum_{i=1}^{L} \hat{\sigma}_i^z )^2$ to the Hamiltonian (\ref{eq.H}). 
 \textcolor{black}{ In Fig.~\ref{fig:globalrange}(a), without global-range interaction, the return probability at RBS $n=3$ decays to 0 for an $L=100$ ring, recognized as the false vacuum decay; while in Fig.~\ref{fig:globalrange}(b), with global-range interaction $\beta=0.2$, the dynamics changes to a persistent oscillation, due primarily to the $\sim\beta$ splitting of the multi-bubble continuum.}
\textcolor{black}{Moreover, the scaling of tunneling splitting, $\Delta E\sim g^{n}L^{1/2}$, captured by Eq.~(\ref{eq.sym-swt-n=3}) for $n=3$, remains valid for global-range interactions, as shown in Fig.~\ref{fig:globalrange}(c). Here, the simulated oscillations for $L=16,50$, and $100$ are plotted in terms of the tunneling splitting $\Delta E=2\pi/T$; the superradiant collective enhancement is again evident.}

\textit{Conclusions.}
In this work we have uncovered a coherent, two-state oscillation phenomenon that arises within the false vacuum decay regime of the TLFIM. Near the resonant condition \(h\approx 2J/n\), the dynamics reduces to an effective two-level problem between \(\lvert\Omega\rangle\) and a symmetric resonant state, characterized by a tunneling splitting \(\Delta E\), whose collective \(\sqrt{L}\) enhancement is captured by a Schrieffer-Wolff treatment in the symmetric subspace. This coherence is further found to persist in large chains when \(n\gtrsim L/2\) due to a bubble-size blockade, or when long-range interactions are present to lift the relevant multi-bubble degeneracies, not confined merely to finite sizes or weak transverse fields. \textcolor{black}{Similar tunneling-splitting-induced oscillations can be realized more generally beyond false vacuum initiation (see SM~\cite{SM} Sec.~VII).} Our understanding suggests inter-domain interaction engineering as a route to controllable recurrences in metastable many-body dynamics \cite{barredo2015coherent}, with potential relevance to metastability \cite{yin2025theory}, spectral-entropy diagnostics \cite{wang2025manybodyquantumgeometrytimedependent},
interaction-based nonlinear interferometric readout \cite{liu2022nonlinear}, etc.

\textit{Acknowledgments.}
This work is supported by the National Natural Science Foundation of China (NSFC Grants No. 92565306 and No. 92265205), and by the Quantum Science and Technology-National Science and Technology Major Project (2021ZD0302104). XW thanks the Gongji Cloud for the support of the computational resources for plotting the phase diagrams, and appreciates helpful discussions with Xin Chen, Laura Batini, and Xinhui Liang.

\bibliography{apssamp}

\clearpage
\onecolumngrid
\setcounter{secnumdepth}{2}  
\pagenumbering{arabic}   
\setcounter{page}{1}

\renewcommand{\thefigure}{S\arabic{figure}}
\setcounter{figure}{0}

\renewcommand{\thesection}{\Roman{section}}
\renewcommand{\thesubsection}{\Roman{section}.\Alph{subsection}}
\renewcommand{\theequation}{S\arabic{equation}}
\setcounter{section}{0}
\setcounter{equation}{0}

\makeatletter
\renewcommand\section{\@startsection{section}{1}{0pt}%
  {24pt plus 6pt minus 6pt}
  {16pt plus 6pt minus 4pt}
  {\centering\normalfont\normalsize\bfseries}} 
\makeatother

\begingroup
\centering
\large\bfseries Supplemental Material \\
Coherent Two-State Oscillations in False Vacuum Decay Regimes\par
\vspace{6pt}
\normalfont\normalsize   
Peiyun Ge$^{1}$, Xiao Wang$^{1,2}$, Yu-Xin Chao$^{1}$, Rong Lu$^{1}$, and Li You$^{1,2,3,4}$\par
\vspace{2pt}
\textit{$^{1}$State Key Laboratory of Low Dimensional Quantum Physics,\\ Department of Physics, Tsinghua University, Beijing 100084, China}\\
\textit{$^{2}$Beijing Academy of Quantum Information Sciences, Beijing 100193, China}\\
\textit{$^{3}$Frontier Science Center for Quantum Information, Beijing 100084, China}\\
\textit{$^{4}$Hefei National Laboratory, Hefei, Anhui 230088, China}\par
\vspace{8pt}
\justifying
In this Supplemental Material, we provide additional derivations and details that support the main results of the Letter. \textcolor{black}{In Section \ref{appendix.sec.tracking} we explain why false vacuum $|\Omega\rangle$ remains close to a single eigenstate in static regimes and transfers its dominant overlap across an avoided crossing.} In Section \ref{appendix.sec.I} we explain the coherent dynamics observed in Fig.~\ref{fig:all} using Eqs.~(\ref{eq.two-state-bloch-oscillation}) and (\ref{eq.S-state}) of the main text. In Section \ref{appendix.SWT}, we construct the SWT in the symmetric subspace and derive the effective two-state Hamiltonian showing a superradiant-like enhancement of the oscillation frequency. In Section \ref{appendix.sec.bubblesize} we \textcolor{black}{show the persistence of bubble size blockade effect in larger systems, and} demonstrate the robustness of the two-state oscillation against the interaction disorder in the blockade-facilitated large-$g$ regime. In Section \ref{appendix.sec.long-range} we show numerically that the coherent two-state oscillation with $r^{-3}$ long-range interactions can still be described by Eq.~(\ref{eq.two-state-bloch-oscillation}). In Section \ref{appendix.sec.global} we show that in large systems, the coherence of the two-state oscillation can be drastically enhanced by cavity-mediated global-range interactions. In Section \ref{appendix.orbital} we analyze the coherent orbital structure arising from the consecutive coherent two-state oscillations starting from the symmetric state $\vert S_n\rangle$.
\par
\endgroup
\vspace{12pt}

\section{Tracking the spectrum with increasing $h$}\label{appendix.sec.tracking}

\textcolor{black}{In our work, the false vacuum state $\vert \Omega\rangle$ is prepared as the ground state of the TLFIM,
\begin{equation}\label{SM.eq.H}
    \hat{H}(g,h) = -J \sum_{i=1}^{L} \hat{\sigma}_i^z \hat{\sigma}_{i+1}^z - g \sum_{i=1}^{L} \hat{\sigma}_i^x + h \sum_{i=1}^{L} \hat{\sigma}_i^z .
\end{equation}
To prepare $|\Omega\rangle$, the longitudinal field $h$ is set to be a small negative value $h_0<0$ that explicitly breaks the $\mathbb{Z}_2$-symmetry and selects a symmetry-broken sector, while remaining weak enough not to appreciably modify the wavefunctions in each sector.
In practice, this is achieved by scanning $h_0$ from $-10^{-9}$ to $-10^{-1}$, and identifying the plateau of the ground state expectation value (e.g., the magnetization $\langle\sigma^z\rangle$) of $\hat{H}(g,h_0)$. 
We then choose $h_0$ within this plateau; for the parameters studied here, representative values are $h_0\simeq-10^{-2}$ for $L\le 20$ and $h_0\simeq-10^{-4}$ for $L=100$.
Starting from $|\Omega\rangle$ prepared above, we quench the $h$ from $h_0$ to a positive value and observe the subsequent dynamics.}

\begin{figure}[h!]
    \centering
    \includegraphics[width=0.8\linewidth]{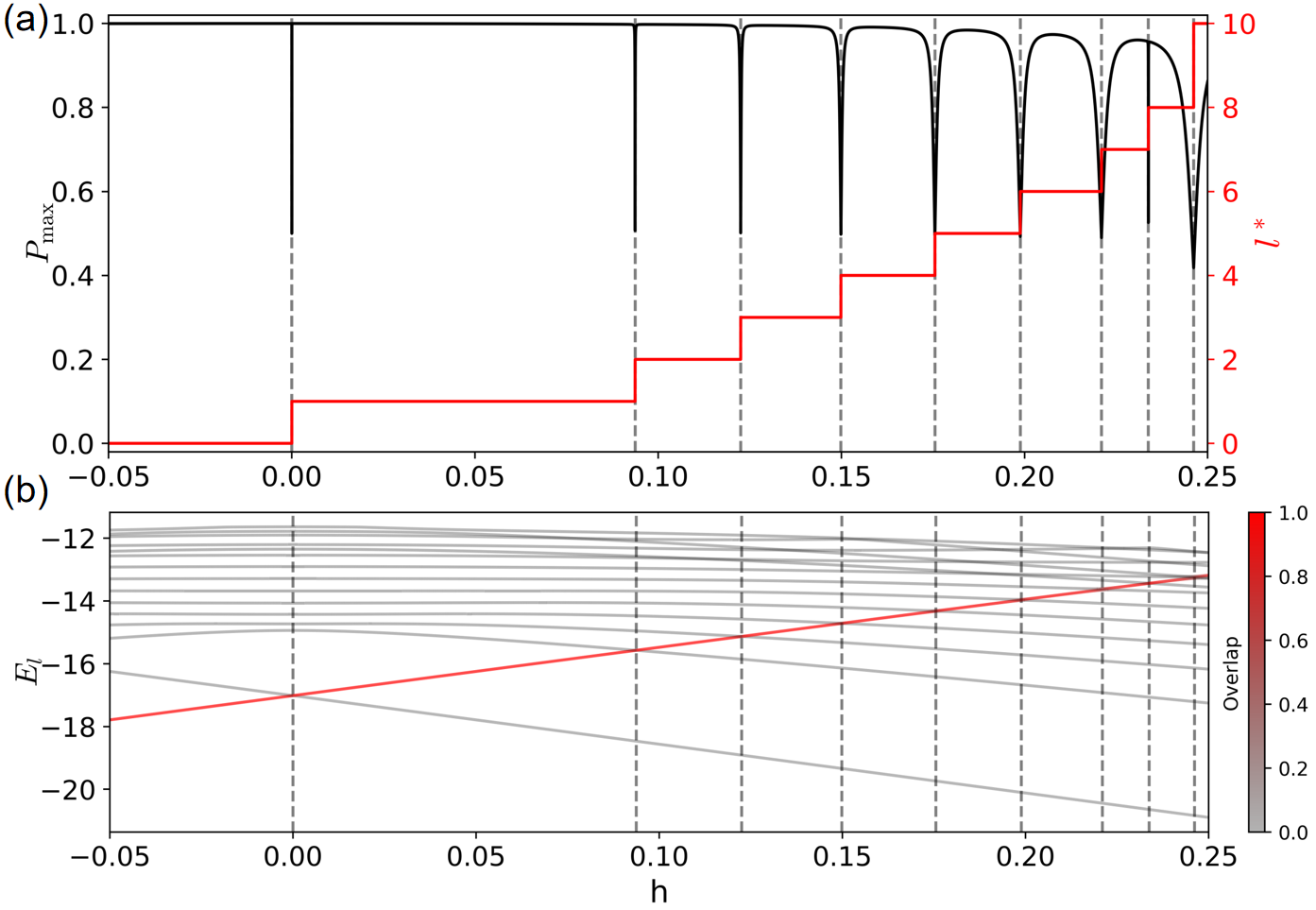}
    \caption{\textcolor{black}{State tracking for an $L=16$ ring at $g=0.5$. (a) The maximal overlap $P_{\max}(h)\equiv \max_l|\langle E_l(h)|\Omega\rangle|^2$ stays close to $1$ in each static regime. Across the avoided crossing, the maximizing index $l^\ast(h)\equiv \arg\max_l|\langle E_l(h)|\Omega\rangle|^2$ shifts from $l$ to $l+1$. (b) The lowest 15 energy levels of the post-quench Hamiltonian. After each (avoided) crossing, the eigenstates hosting the dominant overlap with $|\Omega\rangle$ switches. The colormap encodes the overlap of the eigenstates with the false vacuum, $|\langle E_l|\Omega\rangle|^2$. }}
    \label{figstatetrack}
\end{figure}

\textcolor{black}{As sketched in Fig.~\ref{fig:phasediagram}(d)--(f) of the main text, we observe that, when $h$ scans within each static interval, the dominant overlap between $|\Omega\rangle$ and the eigenstates $|E_l\rangle$ of $\hat{H}(g,h)$ remains close to unity. As $h$ scans through the boundary of a static interval, the dominant overlap transfers to the neighboring level after an avoided crossing (see also Fig.~\ref{fig:phasediagram}(g) in the main text). This behavior is further illustrated in Fig.~\ref{figstatetrack}(a): in the static regime, $|\Omega\rangle$ is close to a single eigenstate with fixed level index; while at an avoided crossing it undergoes population transfer. In Fig.~\ref{figstatetrack}(b), we plot the spectrum of the post-quench Hamiltonian $\hat{H}(g,h)$, with the color bar representing the overlap $|\langle E_l(h)|\Omega\rangle|^2$. Figure~\ref{figstatetrack}(b) directly shows the index switching of the dominant overlap at each avoided crossing in the spectrum. Notably, there is one low-lying spectral branch whose energy increases with $h$, which, according to our main text, is identified as $|E_0^+\rangle$. While other states whose energy decrease with $h$ are categorized into the $\{|E_m^-\rangle\}$ sector.} 

\textcolor{black}{Below, we explain why the dominant overlap stays near unity (which means the overlap $|\langle E_0^+|\Omega\rangle|^2$ in the main text remains close to 1) as $h$ scans through several static regimes. The reason is attributed to the low susceptibility in the static regimes and the switching of dominant overlap near the crossovers.}

\subsection{Static regime: $|\Omega\rangle$ stays close to one eigenstate}

\textcolor{black}{In a static regime (away from avoided crossings), starting from $\hat{H}(h)|E_l(h)\rangle=E_l(h)|E_l(h)\rangle$, we use the standard eigenvector-derivative formula in the parallel-transport gauge $\langle E_n|\partial_h E_n\rangle=0$, which reads
\begin{equation}\label{eq:app_evec_deriv}
\partial_h |E_l\rangle=\sum_{m\neq l}\frac{\langle E_m|\partial_h \hat{H}|E_l\rangle}{E_l-E_m}\,|E_m\rangle .
\end{equation}
}

\textcolor{black}{We define the fidelity susceptibility (quantum metric) as
\begin{equation}\label{eq:app_fidsus}
\chi_l(h)\equiv \langle \partial_h E_l|(1-|E_l\rangle\langle E_l|)|\partial_h E_l\rangle
=\sum_{m\neq l}\frac{|\langle E_m|\partial_h \hat{H}|E_l\rangle|^2}{(E_m-E_l)^2}.
\end{equation}
When the off-diagonal matrix elements $\langle E_m|\partial_h \hat{H}|E_l\rangle$ are tiny and the level $l$ is well separated from the rest of the spectrum, $\chi_l(h)\ll 1$. In our case when the transverse field is not too strong, the off-diagonal matrix elements, given by $\langle E_m|\sum_{i=1}^L \hat{\sigma}_i^z|E_l\rangle$, are tiny due to the weak quantum fluctuation around $\sigma^z$ in the eigenstates. This implies that $|E_l(h)\rangle$ changes only weakly with $h$ in that interval.}

\textcolor{black}{We next relate $\chi_l$ to $P_l(h)=|\langle E_l(h)|\Omega\rangle|^2$. Using $\partial_h P_l=2\,\mathrm{Re}\big[\langle\Omega|E_l\rangle\langle\partial_h E_l|\Omega\rangle\big]$ and noting that the component of $\langle \partial_h E_l|$ parallel to $\langle E_l|$ does not contribute to $\partial_h P_l$ (since $\mathrm{Re}\langle E_l|\partial_h E_l\rangle=0$), we may insert $\hat{Q}_l\equiv 1-|E_l\rangle\langle E_l|$ in $\langle\partial_h E_l|\Omega\rangle$. Then Cauchy--Schwarz inequality gives
\begin{align}\label{eq:app_Pl_bound}
|\partial_h P_l(h)|
&=2\Big|\mathrm{Re}\!\big[\langle\Omega|E_l\rangle\langle\partial_h E_l|\hat{Q}_l|\Omega\rangle\big]\Big|
\le 2|\langle\Omega|E_l\rangle|\;|\langle\partial_h E_l|\hat{Q}_l|\Omega\rangle|
\nonumber\\
&\le 2\sqrt{P_l(h)}\,
\sqrt{\langle \partial_h E_l|\hat{Q}_l|\partial_h E_l\rangle}\,
\sqrt{\langle \Omega|\hat{Q}_l|\Omega\rangle}
=2\sqrt{\chi_l(h)}\,\sqrt{P_l(h)\,[1-P_l(h)]}.
\end{align}
Therefore, if $P_l(h_0)=1-\varepsilon$ at some $h_0$ in the static regime and $\chi_l(h)$ stays small throughout that interval, then $P_l(h)$ remains close to unity for all $h$ in that regime. Physically, $|\Omega\rangle$ is ``stuck'' to a single eigenstate because both the coupling $\langle E_m|\partial_h \hat{H}|E_l\rangle$ and the energy-denominator $(E_m-E_l)^{-1}$ suppress the mixing with other levels in Eq.~(\ref{eq:app_evec_deriv}).}

\subsection{Avoided crossing: transfer between $l$ and $l+1$}
\textcolor{black}{Near an avoided crossing between levels $l$ and $l+1$, the energy denominator $|E_{l+1}-E_l|$ becomes small, so the susceptibility \eqref{eq:app_fidsus} peaks and $|\Omega\rangle$ resides mainly in $\{|E_l\rangle,|E_{l+1}\rangle\}$, which is the subspace spanned by the corresponding two eigenstates. When other levels remain well separated, the dynamics and overlaps could be captured by an effective two-level model constructed in a diabatic basis $\{|a\rangle,|b\rangle\}$. Here $|a\rangle$ and $|b\rangle$ reside in two $\mathbb{Z}_2$-symmetry-broken sectors, $\{|E^+_m\rangle\}$ and $\{|E^-_m\rangle\}$, respectively. The effective Hamiltonian in this basis reads
\begin{equation}\label{eq:app_2lvl}
\hat{H}_{\rm eff}(h)=\bar E(h)+\frac{\Delta(h)}{2}\hat{\sigma}_z + V_{\rm eff}(h)\hat{\sigma}_x,
\end{equation}
where $\Delta(h)$ denotes the energy splitting of $|a\rangle$ and $|b\rangle$ and $V_{\rm eff}(h)$ the effective coupling between them. See Sec.~\ref{appendix.SWT} for an example of an explicit derivation of these parameters.}

\textcolor{black}{The adiabatic eigenstates of the Eq.~(\ref{eq:app_2lvl}) are
\begin{align}
|E_l(h)\rangle     &= \cos\theta(h)\,|a\rangle-\sin\theta(h)\,|b\rangle,\\
|E_{l+1}(h)\rangle &= \sin\theta(h)\,|a\rangle+\cos\theta(h)\,|b\rangle,
\end{align}
where
\begin{equation}
\tan\!\big(2\theta(h)\big)=\frac{2|V_{\rm eff}(h)|}{\Delta(h)}.
\end{equation}
If $|\Omega\rangle$ coincides with one diabatic basis, $|\Omega\rangle\simeq |a\rangle$, then
\begin{equation}\label{eq:app_transfer}
|\langle E_l(h)|\Omega\rangle|^2\simeq \cos^2\theta(h),\qquad
|\langle E_{l+1}(h)|\Omega\rangle|^2\simeq \sin^2\theta(h).
\end{equation}
As $h$ scans across the avoided crossing, $\Delta(h)$ changes sign, thus $\theta(h)$ varies by approximately $\pi/2$, so the dominant overlap transfers from the $l$-th eigenstate to the $(l+1)$-th eigenstate. The combination of Secs.~\ref{appendix.sec.tracking}A and \ref{appendix.sec.tracking}B explains why $|\Omega\rangle$ closely follows a single eigenstate within each static regime, while its associated level index switches upon traversing the avoided-crossing region.}

\clearpage
\section{Explanation of the observed coherent dynamics in Fig.~\ref{fig:all}}\label{appendix.sec.I}
In the regime of $h\sim\frac{2J}{n}$ and small $g$, the false vacuum state is expected to be resonant with the symmetric single-bubble state with bubble size $n$, defined as $\vert S_n\rangle=\frac{1}{\sqrt{L}}\sum_i\hat{S}^-_{i}\hat{S}^-_{i+1}...\hat{S}^-_{i+n-1}|\Omega\rangle$ where $\vert \Omega\rangle \approx \vert \uparrow,\uparrow,...,\uparrow\rangle$. [For example, see Eq.~(\ref{eq.S-state}) in the main text for $\vert S_3\rangle$.] In Fig.~\ref{fig:all} in the main text, when the initial false vacuum state evolves into $\vert S_n\rangle$, the magnetization approaches $m(L,n)=\frac{L-2n}{L}$ and the two-site correlation $\langle Z_0Z_r\rangle$ can be expressed as $\langle Z_0Z_r\rangle=\frac{L-2N(L,n,r)}{L}$,
 where $N(L,n,r)$ denotes the number of spin pairs with opposite directions separated by distance $r$,
\begin{equation}\label{eq.oppo-spin-pairs}
N(L,n,r)=
\begin{cases}
2r, & 1\leq r < k, \\
2k, & k\leq r \leq \frac{L}{2}. \\
\end{cases}
\end{equation}
where $k=min\left(n,L - n\right),1\leq r\leq L/2$.
For example, when $L=8$ and $n=3$, one finds $N(r=1)=2,N(r=2)=4,N(r=3)=6,N(r=4)=6$, so $\langle Z_0Z_1\rangle= \frac{1}{2},\langle Z_0Z_2\rangle=0,\langle Z_0Z_3\rangle=-\frac{1}{2},\langle Z_0Z_4\rangle=-\frac{1}{2}$, which are in quantitative agreement with the correlations shown in Fig.~\ref{fig:all} in the main text.

\section{The effective two-state Hamiltonian given by SWT}\label{appendix.SWT}
We separate the Hamiltonian (\ref{SM.eq.H}) into $\hat{H}=\hat{H}_0 + \hat{H}^{(1)}$, where $\hat{H}^{(1)}$ represents the transverse field. Near the $n=3$ resonance, we restrict $\hat{H}$ to a subspace spanned by 5 symmetric states, $\{$ $|\Omega\rangle=|\uparrow\uparrow...\rangle$, $|S_1\rangle=\frac{1}{\sqrt{L}}\sum_i\hat{S}^-_{i}|\Omega\rangle$, $|S_2\rangle=\frac{1}{\sqrt{L}}\sum_i\hat{S}^-_{i}\hat{S}^-_{i+1}|\Omega\rangle$, $|S_2'\rangle=\frac{1}{\sqrt{L}}\sum_i\hat{S}^-_{i}\hat{S}^-_{i+2}|\Omega\rangle$, $|S_3\rangle=\frac{1}{\sqrt{L}}\sum_i\hat{S}^-_{i}\hat{S}^-_{i+1}\hat{S}^-_{i+2}|\Omega\rangle$$\}$ as shown in Fig.~\ref{fig:all} in the main text. In this subspace, the unperturbed term becomes a diagonal matrix,
\begin{equation}
    \hat{H}_0 = 
\begin{pmatrix}
-L+Lh &  &  &  &  \\
 & 4-L+(L-2)h &  &  &  \\
 &  & 4-L+(L-4)h &  &  \\
 &  &  & 8-L+(L-4)h &  \\
 &  &  &  & 4-L+(L-6)h
\end{pmatrix},
\end{equation}
and the perturbation term reads 
\begin{equation}
    \hat{H}^{(1)} = -g
\begin{pmatrix}
0 & \sqrt{L} & 0 & 0 & 0 \\
\sqrt{L} & 0 & 2 & 2 & 0 \\
0 & 2 & 0 & 0 & 2 \\
0 & 2 & 0 & 0 & 1 \\
0 & 0 & 2 & 1 & 0
\end{pmatrix}.
\end{equation}
Near $n=3$ resonance, a block-diagonal structure emerges from the degeneracy of $\hat{H}_0$, where $\vert \Omega\rangle$ and $\vert S_3\rangle$ span a degenerate block.
In the Schrieffer-Wolff transformation, 
\begin{equation}
\begin{split}
\hat{H}' =  e^{\hat{S}} \hat{H} e^{-\hat{S}}
&= \hat{H}_0 + \hat{H}^{(1)} + [\hat{S},\hat{H}_0 + \hat{H}^{(1)}] \\
&~~+ \frac{1}{2!} [\hat{S},[\hat{S},\hat{H}_0 + \hat{H}^{(1)}]] \\
&~~+ \frac{1}{3!} [\hat{S},[\hat{S},[\hat{S},\hat{H}_0 + \hat{H}^{(1)}]]] + ...,
\end{split}
\end{equation}
the anti-Hermitian operator $\hat{S}$ is chosen to eliminate the off-block-diagonal term in the transformed Hamiltonian $\hat{H}'$. We expand $\hat{S}$ in orders of $g$, $\hat{S}=\sum_{i=n}^{\infty} \hat{S}^{(n)}$, such that $\hat{S}^{(i)}$ eliminates the $O(g^n)$ off-block-diagonal term in $\hat{H}'$. By eliminating the off-block-diagonal terms to $O(g)$ and then $O(g^2)$ order, we respectively obtain $\hat{S}^{(1)}$ and $\hat{S}^{(2)}$, i.e.,
\begin{equation}
    \hat{S}^{(1)} = -g
\begin{pmatrix}
0 & \frac{\sqrt{L} }{2(h - 2)} & 0 & 0 & 0 \\
-\frac{\sqrt{L} }{2h - 4} & 0 & \frac{1}{h} & \frac{1}{h - 2} & 0 \\
0 & -\frac{1}{h} & 0 & 0 & \frac{1}{h} \\
0 & -\frac{1}{h - 2} & 0 & 0 & \frac{1}{2(h + 2)} \\
0 & 0 & -\frac{1}{h} & -\frac{1}{2h + 4} & 0
\end{pmatrix},
\end{equation}
\begin{equation}
    \hat{S}^{(2)} = g^2 
\begin{pmatrix}
0 & 0 & \frac{4 \sqrt{L}}{h (16 h^{2} - 48 h + 32)} & 0 & 0 \\
0 & 0 & 0 & 0 & \frac{2}{h (4 h^{2} - 16)} \\
-\frac{4 \sqrt{L}}{h (16 h^{2} - 48 h + 32)} & 0 & 0 & \frac{16 h^{2} + 48 h - 32}{h (64 h^{2} - 256)} & 0 \\
0 & 0 & \frac{-16 h^{2} - 48 h + 32}{h (64 h^{2} - 256)} & 0 & 0 \\
0 & -\frac{2}{h (4 h^{2} - 16)} & 0 & 0 & 0
\end{pmatrix}.
\end{equation}
The effective two-state Hamiltonian, correct to $O(g^3)$, is then extracted from the degenerate block, which reads
\begin{equation}\label{eq.appendix.sym-swt-n=3}
    \hat{H}_{eff}^{mod}=\begin{pmatrix}
    \langle \Omega\vert \hat{H}'\vert \Omega \rangle & \langle \Omega\vert \hat{H}'\vert S_3 \rangle\\
    \langle S_3\vert \hat{H}'\vert \Omega \rangle & \langle S_3\vert \hat{H}'\vert S_3 \rangle
    \end{pmatrix}
    =\begin{pmatrix}
\frac{ L g^{2}}{2 h - 4}+L \left(h - 1\right) & -\frac{ \left( h \left(9 h - 10 \right)-8\right) g^{3}\sqrt{L}}{12 \left(h-2 \right)^{2} h^{2} \left( h^{2}+h-2\right)} \\
-\frac{ \left( h \left(9 h - 10 \right)-8\right) g^{3}\sqrt{L}}{12 \left(h-2 \right)^{2} h^{2} \left( h^{2}+h-2\right)} & -\frac{g^2 (5h + 8)}{2h (h + 2)} - L + h \left(L - 6\right) + 4 
\end{pmatrix}
+ O(g^4).
\end{equation}
In Eq.~(\ref{eq.appendix.sym-swt-n=3}), the $O(g^0)$ term represents the energy detuning in $\hat{H}_0$. The $O(g^2)$ diagonal self-energy term comes from the second-order virtual process, akin to the AC Stark shift. The $O(g^3)$ term serves as the Rabi frequency of the effective two-state dynamics, and the $\sqrt{L}$ enhancement comes from the normalization factor of the symmetric state. The main text's Eq.~(\ref{eq.sym-swt-n=3}) is recovered from Eq.~(\ref{eq.appendix.sym-swt-n=3}) by defining
\begin{equation}
\begin{split}
\kappa(h) &= \frac{  h \left(9 h - 10 \right)-8 }{12 \left(h-2 \right)^{2} h^{2} \left( h^{2}+h-2\right)}, \\
E_0(h,g)&=-\frac{g^2 (5h + 8)}{2h (h + 2)}+L \left(h - 1\right),\\
\Delta(h,g,L)&=g^2\left(\frac{ L} {2 h - 4}+\frac{ (5h + 8)}{2h (h + 2)}\right).
\end{split}
\end{equation}

\section{Scaling and robustness of the bubble size blockade}\label{appendix.sec.bubblesize}
\textcolor{black}{In this section, we provide numerical evidence that the bubble-size blockade extends beyond the exact-diagonalization regime and is robust against interaction disorder in large-$g$ regime. In particular, we focus on the resonance window at RBS order $n=L-1$, where the bubble-size blockade effect is strong.}

\subsection{Bubble-size blockade at $n=L-1$: scaling to larger systems}

\begin{figure}[b!]
    \centering
  \includegraphics[width=1\textwidth]{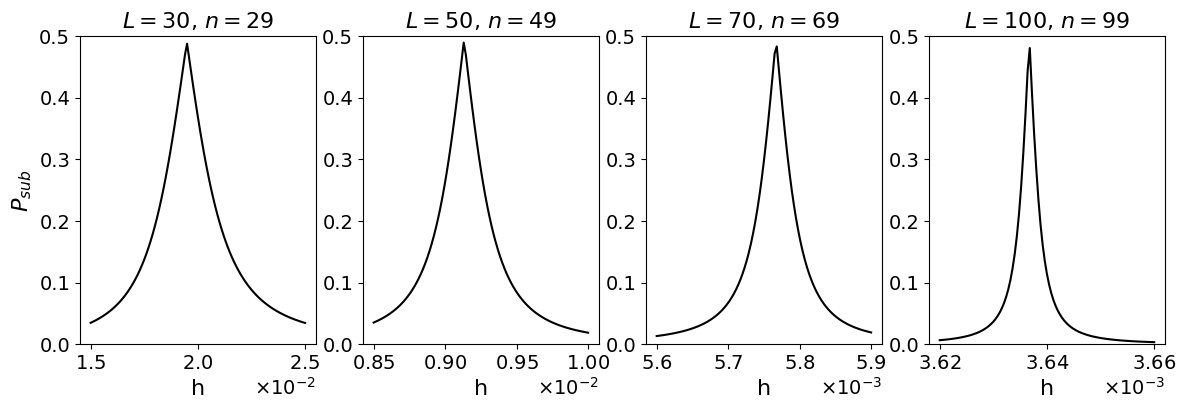} 
  \caption{\textcolor{black}{Sub-leading overlap $P_{\rm sub}$ at the RBS order $n=L-1$ for $L=30,50,70$, and $100$ at $g=0.9$. For all system sizes, $P_{\rm sub}$ approaches $0.5$ at resonance, indicating the effective two-level hybridization between $|\Omega\rangle$ and the resonant bubble state $|E_1^-\rangle$. Small deviations from $0.5$ arise primarily from the finite sampling resolution in $h$.}}
  \label{fig:BLOCKADE_L_scaling}
\end{figure}

\begin{figure}[h!]
    \centering
  \includegraphics[width=0.6\textwidth]{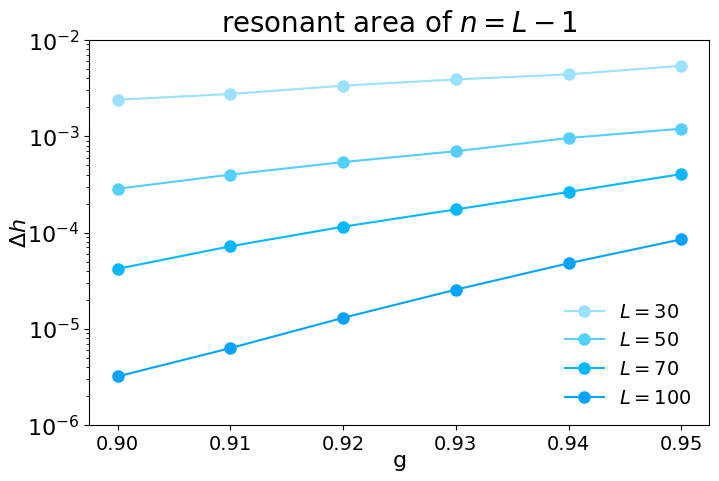} 
  \caption{\textcolor{black}{The resonant window $\Delta h$ at the $n=L-1$ (RBS) blockade condition broadens with increasing $g$. Here, $\Delta h$ is defined according to $P_{\rm sub}(h)>0.25$. Shown is $\Delta h$ for ring-size $L=30,50,70$, and $100$ and transverse fields $g\in[0.90,0.95]$. While $\Delta h$ decreases with increasing $L$, it grows rapidly with $g$ (approximately exponentially over the range shown), partially compensating the size-induced narrowing of the $n=L-1$ resonance.}}
  \label{fig:BLOCKADE_g_broadening}
\end{figure}

\textcolor{black}{The bubble-size blockade effect discussed in the main text is not restricted to small system sizes. To assess its persistence in larger rings, we perform MPS-based DMRG calculations with bond dimension $D=100$ for $L=30,50,70$, and $100$, and evaluate the overlaps between the false vacuum state $|\Omega\rangle$ and the low-lying eigenstates of $\hat{H}$ in Eq.~(\ref{SM.eq.H}). As shown in Fig.~\ref{fig:BLOCKADE_L_scaling}, the sub-leading overlap $P_{\rm sub}$ at the $n=L-1$ resonance approaches $0.5$ for all simulated $L$, indicating a robust near-resonant hybridization between $|\Omega\rangle$ and the corresponding resonant bubble configuration $|E_1^-\rangle$ even in large systems. This behavior verifies the mechanism sketched in Fig.~\ref{fig:phasediagram}(e) of the main text.}

\textcolor{black}{As the system size increases, the parameter region for two-state oscillation becomes harder to find: the resonant regime, as a function of the longitudinal field $h$, becomes exponentially narrower. In other words, the characteristic width $\Delta h$ of the region in which $P_{\rm sub}$ is appreciable decreases with $L$. However, this narrowing issue can be mitigated by increasing the transverse field $g$, which broadens the effective resonance window $\Delta h$, as illustrated in Fig.~\ref{fig:BLOCKADE_g_broadening}. Together, these results demonstrate that the blockade physics and the associated two-level structure remain observable beyond the exact-diagonalization regime, and can be made more experimentally accessible by increasing $g$.}

\subsection{Robustness against disorder interaction at large $g$}\label{appendix.sec.disorder}
\begin{figure}[h!]
    \centering
  \includegraphics[width=0.9\textwidth]{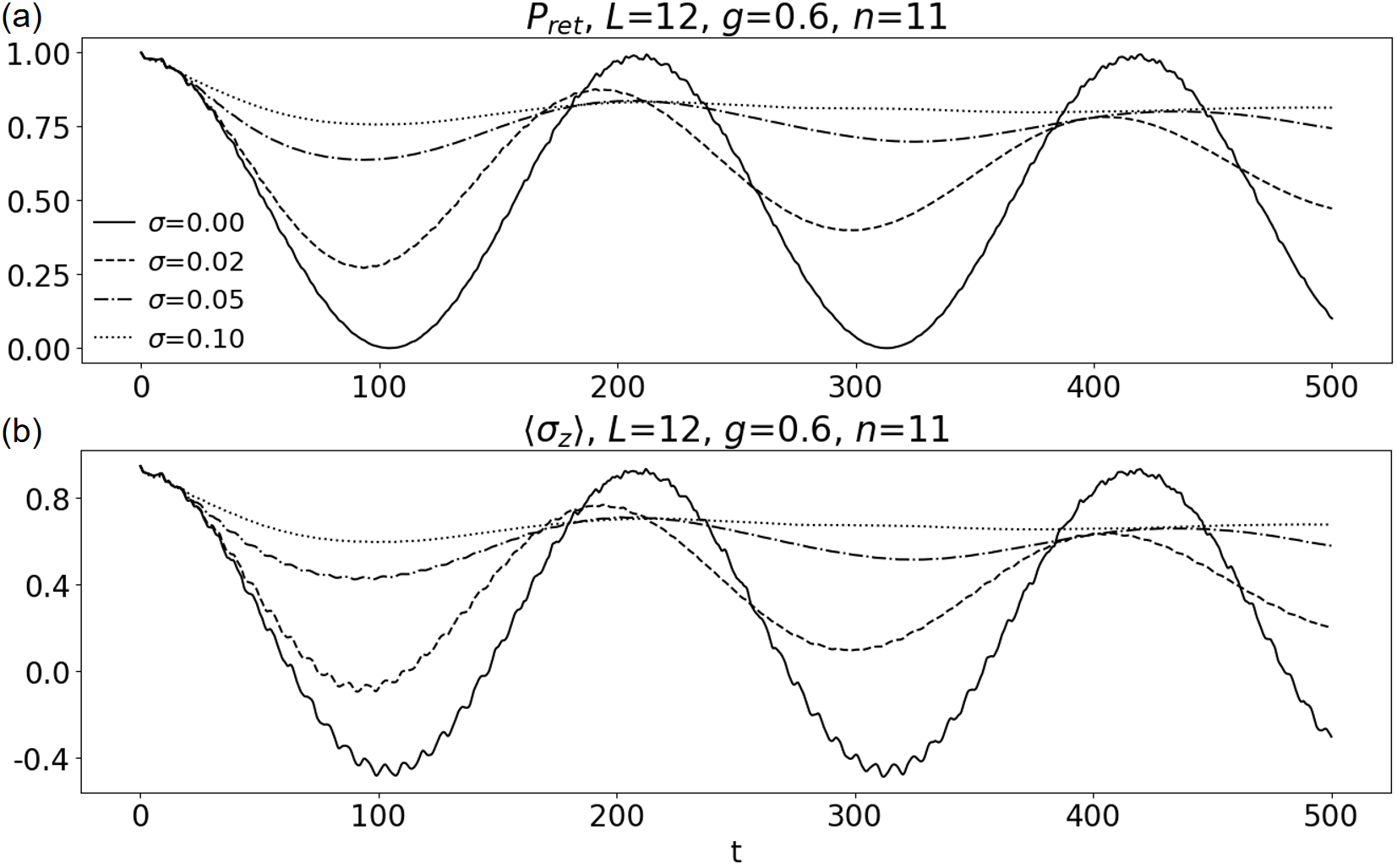} 
  \caption{
  Dynamics of an $L=12$ ring at $g=0.6$, evaluated at the RBS order $n=11$ (around $h=0.1223$), for varying disorder strength $\sigma$. (a) Loschmidt echo, averaged over $1000$ disorder realizations for each $\sigma$. (b) Magnetization.}
  \label{fig11}
\end{figure}

\begin{figure}[h!]
    \centering
  \includegraphics[width=0.9\textwidth]{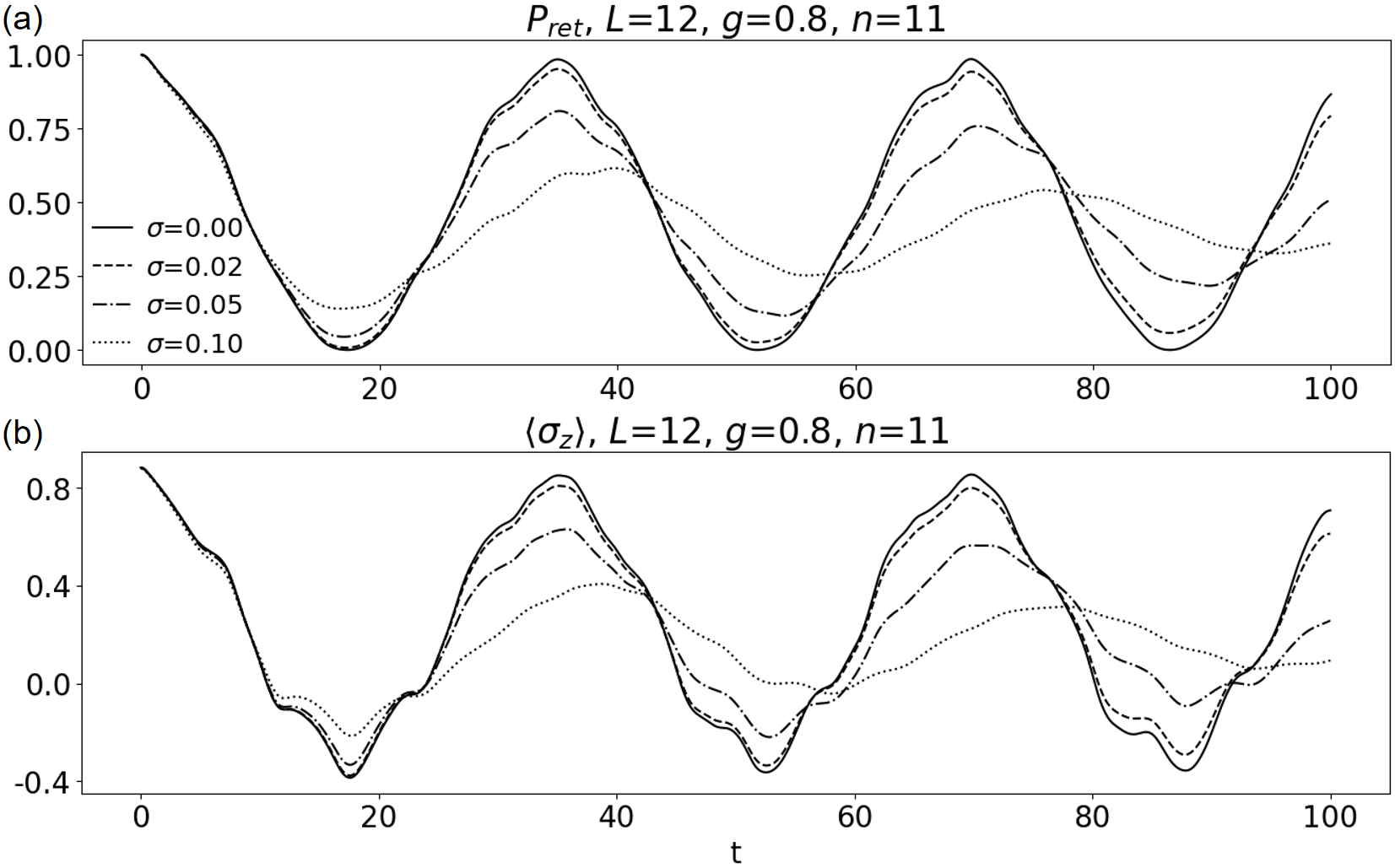} 
  \caption{Same as Fig.~\ref{fig11} but with transverse field increased to $g=0.8$. At this value of $g$, the $n=11$ resonance occurs at $h=0.09875$.}
  \label{fig10}
\end{figure}

In real experiments, fluctuations in the spin-spin distance lead to a modified Hamiltonian, 
\begin{equation}
\hat{H} = - \sum_{i=1}^{L} J_i\hat{\sigma}_i^z \hat{\sigma}_{i+1}^z - g \sum_{i=1}^{L} \hat{\sigma}_i^x + h \sum_{i=1}^{L} \hat{\sigma}_i^z,
\end{equation}
where the couplings $J_i$ are modeled as Gaussian variables centered at $J$ with standard deviation $\sigma$. For this Hamiltonian, we compute the disorder-averaged dynamics over $1000$ realizations, which provides experimentally relevant expectation values. As shown in Fig.~\ref{fig11}, at the $n=11$ resonance with a relatively weak transverse field $g=0.6$, the Loschmidt echo and magnetization are highly sensitive to disorder, such that for deviation $\sigma=0.02$ the coherent oscillations are substantially suppressed. In contrast, Fig.~\ref{fig10} demonstrates that for a larger transverse field $g=0.8$, the dynamics at the $n=11$ resonance remain nearly unchanged until the disorder strength exceeds $\sigma \gtrsim 0.05$. 

\clearpage
\section{The coherent dynamics under $r^{-3}$ interaction}\label{appendix.sec.long-range}

\begin{figure}[th]
  \centering
  \includegraphics[width=0.8\textwidth]{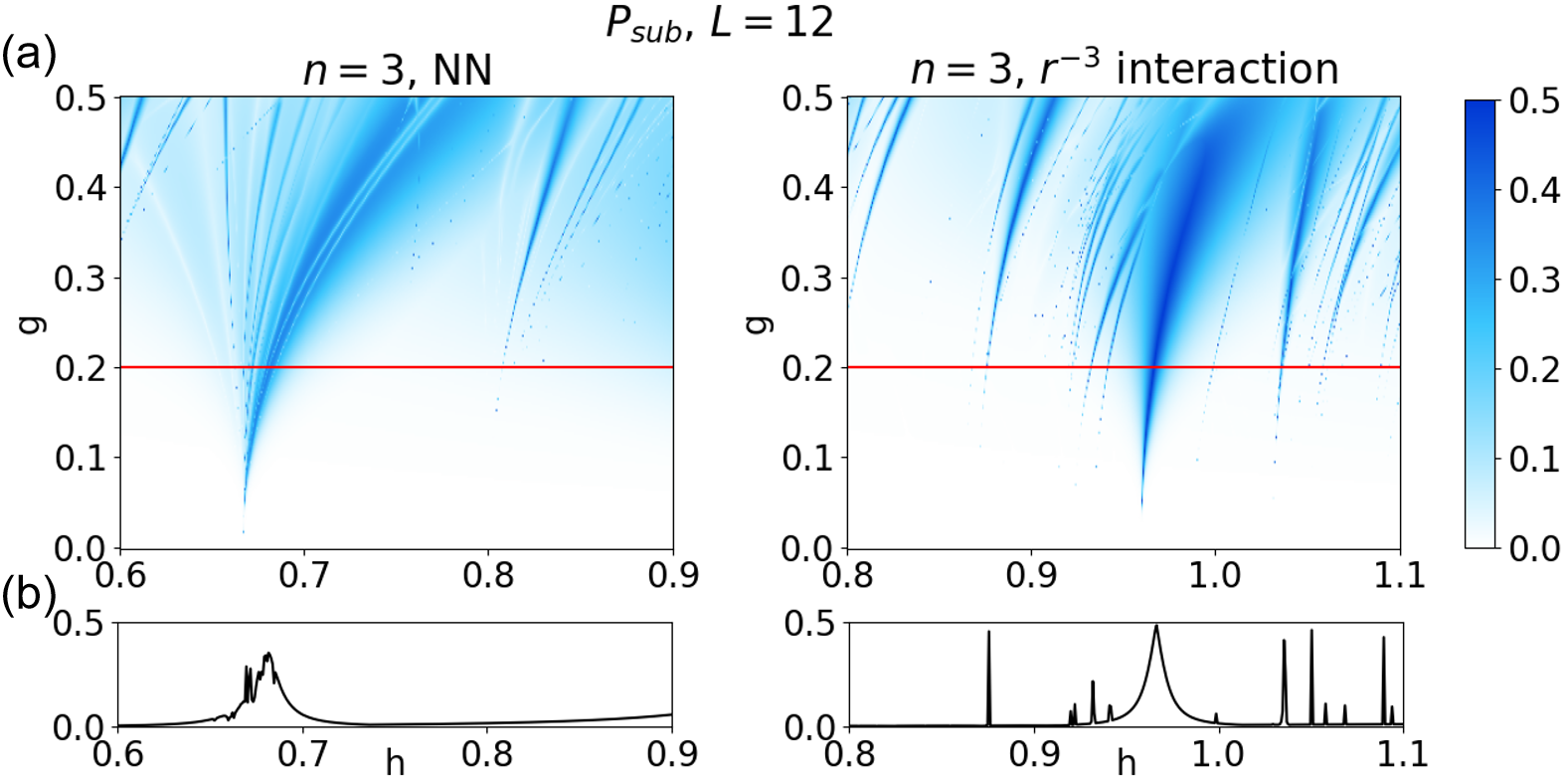} 

  \caption{(a) The sub-leading overlap between the false vacuum state and the eigenstates of $\hat{H}$. The left panel corresponds to the NN Ising model in Eq.~(\ref{SM.eq.H}). The right panel shows the model with $r^{-3}$ type interactions. (b) Horizontal cut of (a) at $g=0.2$ for $L=12$.}
  \label{fig-alpha3}
\end{figure}

\textcolor{black}{As shown in Fig.~\ref{fig-alpha3}, for an $L=12$ ring with $r^{-3}$ long-range Ising interactions (i.e., with the Ising term modified into $-J \sum_{i<j} |i - j|^{-3} \hat{\sigma}_i^z \hat{\sigma}_j^z$  \cite{jurcevic2014quasiparticle,zhang2017observation}), the sub-leading overlap $P_{\text{sub}}$ around RBS order $n=3$ can again reach 0.5. Moreover, the dynamics of two-state oscillation can still be approximated by Eqs. (\ref{eq.two-state-bloch-oscillation}) and (\ref{eq.sym-swt-n=3}) in the main text, with the parameters $\kappa$ and $\Delta$ renormalized by the long range part of the interaction.}

\begin{figure}[th!]
    \centering
  \includegraphics[width=0.7\textwidth]{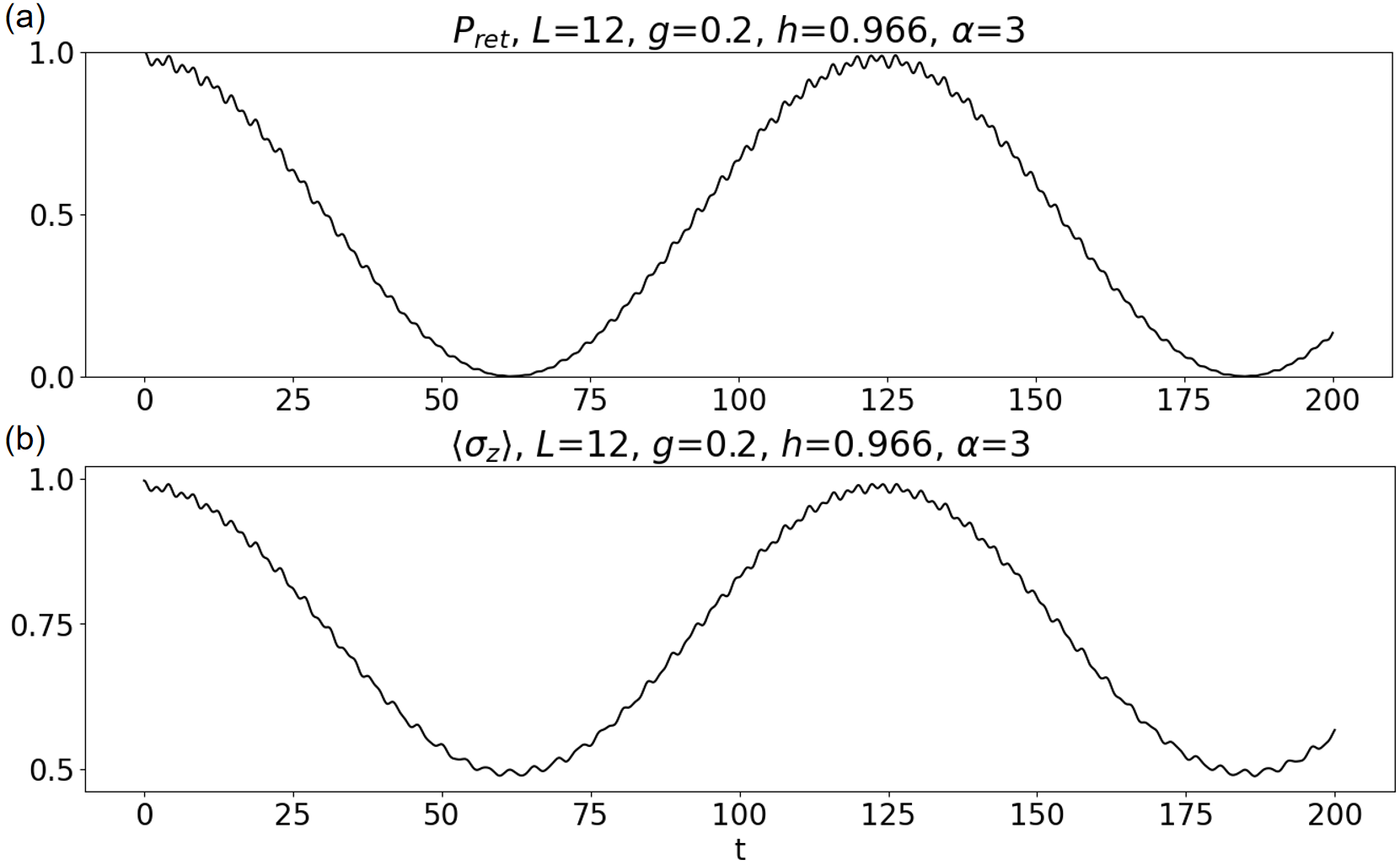} 
  \caption{The coherent dynamics of an $L=12$ ring with $r^{-3}$ interaction. (a) The return probability of $\vert\Omega\rangle$ at $g=0.2$ and $h=0.966$, showing the two-state oscillation. (b) The magnetization of the system.}
  \label{fig-append-alpha3}
\end{figure}

In Fig.~\ref{fig-append-alpha3}, for an $L=12$ ring, we plot the dynamics at a parameter point, $g=0.2$, $h=0.965$, lying in the blue regime in the right panel of Fig.~\ref{fig-alpha3}(a). During the oscillation, the magnetization $\langle \hat{\sigma}^z \rangle$ reaches a minimum of $0.5$, which fits well with the relation $m(L,n)=\frac{L-2n}{L}$ in Sec.~\ref{appendix.sec.I} for $n=3$.

\section{Adding a global-range spin-squeezing interaction}\label{appendix.sec.global}

We off-resonantly couple the many-spin system to a cavity, with the following longitudinal qubit-oscillator interaction,
\begin{equation}
\begin{split}
\hat{H} &= -J \sum_{i=1}^{L} \hat{\sigma}_i^z \hat{\sigma}_{i+1}^z - g \sum_{i=1}^{L} \hat{\sigma}_i^x + h \sum_{i=1}^{L} \hat{\sigma}_i^z \\
& + \omega_c \hat{a}^{\dagger}\hat{a} + g_c(\hat{a} + \hat{a}^{\dagger}) \left( \sum_{i=1}^{L} \hat{\sigma}_i^z \right).
\end{split}
\end{equation}
By adiabatically eliminating the cavity mode $\hat{a}$, we can engineer a global-range spin-squeezing term (with strength $\beta$) in the TLFIM Hamiltonian,
\begin{equation}\label{eq.appendix.Hcav-med-int}
\begin{split}
\hat{H} &= -J \sum_{i=1}^{L} \hat{\sigma}_i^z \hat{\sigma}_{i+1}^z - g \sum_{i=1}^{L} \hat{\sigma}_i^x + h \sum_{i=1}^{L} \hat{\sigma}_i^z  - \frac{\beta J}{2L} \left( \sum_{i=1}^{L} \hat{\sigma}_i^z \right)^2.
\end{split}
\end{equation}
\begin{figure}[ht!]
    \centering
    \includegraphics[width=0.8\linewidth]{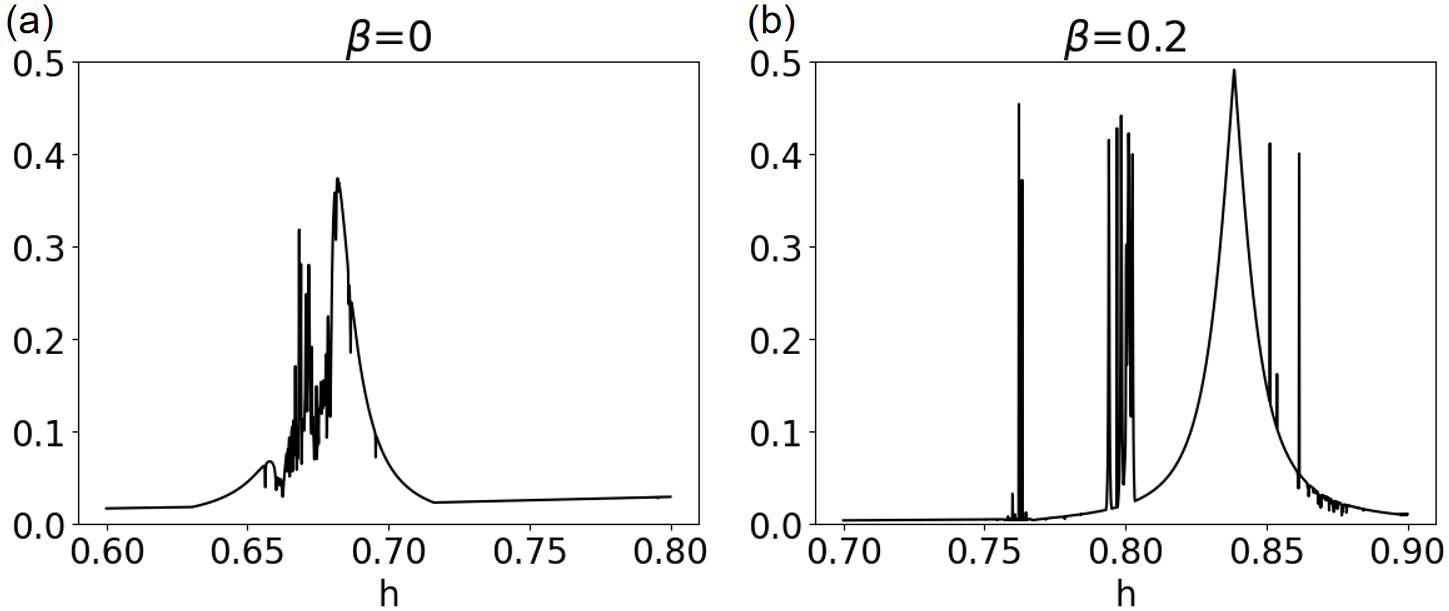}
    \caption{The sub-leading overlap, $P_{\text{sub}}$, of state $\vert \Omega\rangle$ in an $L=16$ ring around RBS order $n=3$ with $g=0.2$ and global-range interaction strength $\beta$. (a) $\beta=0$, corresponding to the nearest-neighbour Ising model. (b) $\beta=0.2$, the degeneracy between the single-bubble and multi-bubble manifolds is lifted. }
    \label{fig:cavity}
\end{figure}

This cavity-mediated interaction favors the ferromagnetism. When the chain length is large compared to the RBS order, such that $L>4n$, a two-bubble state will exhibit a smaller magnetization strength compared to a single-bubble state, and thus a larger energy according to Eq. (\ref{eq.appendix.Hcav-med-int}).
\textcolor{black}{In the small $g$ limit, considering the multi-bubble state containing $k$ size-$n$-bubbles, this global interaction leads to an additional energy gap with respect to the false vacuum
\begin{equation}
    -\frac{\beta J}{2L}[(L-2nk)^2-L^2]=-\frac{\beta J}{2L}(4n^2k^2-4nkL)=2\beta Jnk\left(1-\frac{nk}{L}\right).
\end{equation}
This gap lifts the degeneracy of the multi-bubble continuum, splitting the energy between states with different $k$. 
The splitting is proportional to $\beta$, and becomes independent of system size $L$ when $nk\ll L$.
In Fig.~\ref{fig:cavity}, we plot the sub-leading overlap $P_{\text{sub}}$ of the state $|\Omega\rangle$ in an $L=16$ ring, which shows that $P_{\text{sub}}$ can be restored to about $0.5$ with global-range interaction. This global-range interaction pushes the forest of multi-bubble resonances outside the single-bubble resonance, thereby enhancing the coherence of the oscillation between $\lvert\Omega\rangle$ and $\lvert S_3\rangle$. The shift of the resonant $h$ for single-bubble state is about $h_{res}(\beta)-h_{res}(0)=\beta J\left(1-\frac{n}{L}\right)$.}

\textcolor{black}{For larger $L$, the global-range interaction continues to stabilize persistent two-state oscillations due to the splitting of the multi-bubble continuum, as shown in Fig.~\ref{fig:globalrange} of the main text. Throughout these simulations, the MPS bond dimension is capped at $D=20$; convergence is verified by comparison with simulations at $D=100$.}

\section{Coherent generation of multi-bubble states from $\vert S_3\rangle$} \label{appendix.orbital}
\textcolor{black}{We have shown that the false vacuum state $\vert\Omega\rangle$ can coherently evolve into various single-bubble configurations, as studied in the main text. Similar dynamics can also occur from other initial states: appropriately chosen states can couple strongly to a specific multi-bubble manifold and exhibit near-resonant oscillations rather than decay. To illustrate this mechanism, we consider the post-quench dynamics starting from the symmetric state $\vert S_3\rangle$.}

\begin{figure}[h!]
    \centering
    \includegraphics[width=1.0\linewidth]{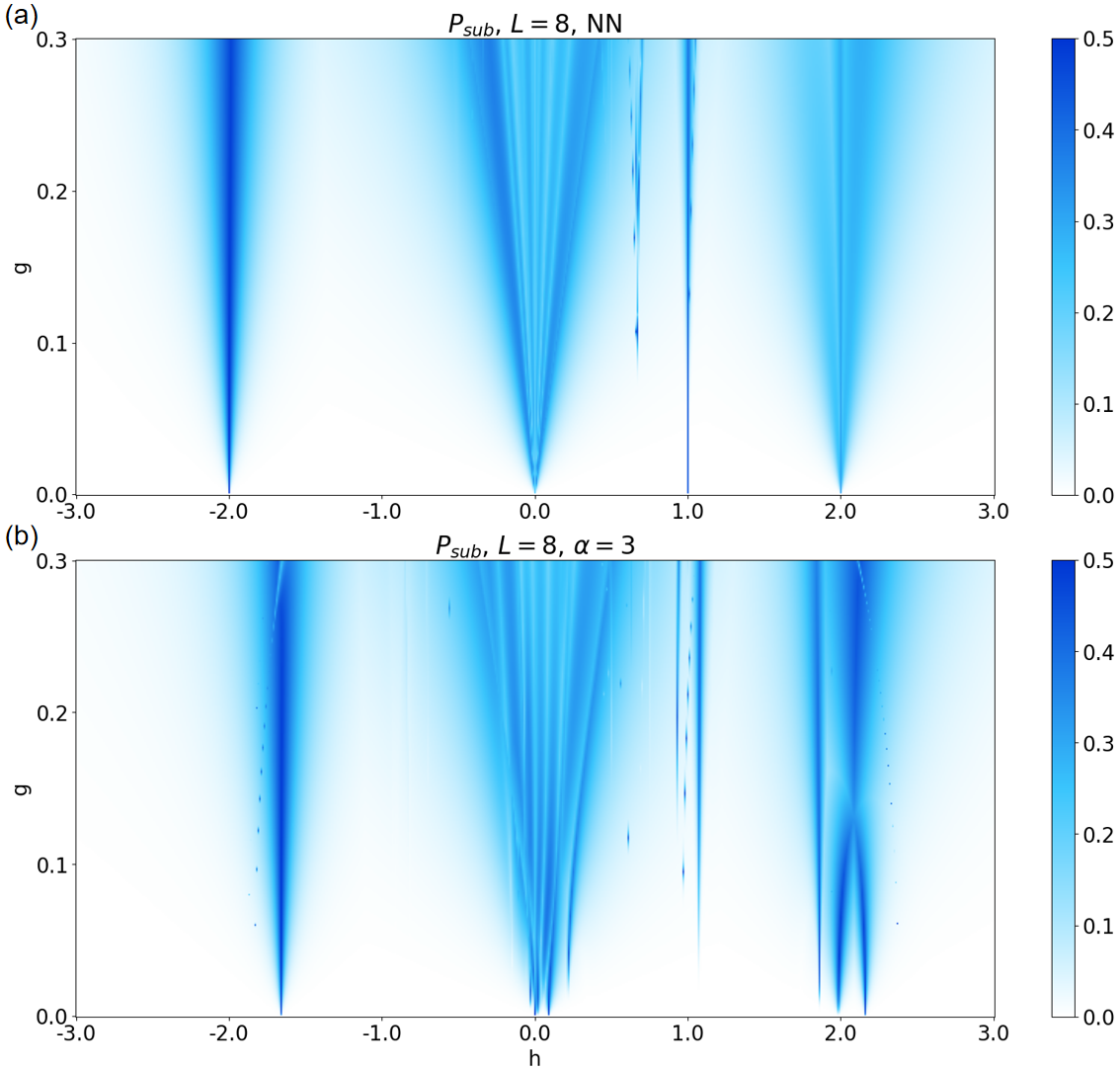}
    \caption{The sub-leading overlap between $\vert S_3 \rangle$ and the eigenstates of the Hamiltonian when $L=8$. (a) The case for the NN Hamiltonian in Eq. (\ref{SM.eq.H}). (b) The same as (a) but with the Ising term replaced by a long-range interaction $-J \sum_{i<j} |i - j|^{-3} \hat{\sigma}_i^z \hat{\sigma}_j^z$.}
    \label{S3subleading}
\end{figure}

Starting from the single-bubble state $\vert S_3\rangle$ defined in Eq.~(\ref{eq.S-state}) of the main text, we find that by modifying the field strengths $h$ and $g$ of the post-quench Hamiltonian $\hat{H}(g,h)$ in Eq.~(\ref{SM.eq.H}), several coherent paths emerge in the subsequent dynamics, which host two-state oscillations between $\vert S_3\rangle$ and various multi-bubble states. These paths are shown in Fig.~\ref{S3subleading}(a), where we plot the sub-leading overlap $P_{\text{sub}}$ between $\vert S_3\rangle$ and the eigenstates of $\hat{H}(g,h)$ for an $L=8$ ring. The blue vertical line at $h=-2$ reflects the resonance between $\vert S_3\rangle$ and $\vert S_{101}\rangle = \frac{1}{\sqrt{L}} \sum\limits_{i} \hat{S}^-_{i-1} \hat{S}^-_{i+1} \vert\Omega\rangle$, while the lines at $h=1$ and $h=2/3$ reveal the resonances $\vert S_3\rangle \leftrightarrow \frac{1}{\sqrt{2L}} \sum\limits_{i} \hat{S}^-_{i-1} \hat{S}^-_{i} \hat{S}^-_{i+1} (\hat{S}^-_{i+3} \hat{S}^-_{i+4} + \hat{S}^-_{i-3} \hat{S}^-_{i-4} )   \vert\Omega\rangle$ and $\vert S_3\rangle \leftrightarrow \vert \Omega\rangle$, respectively. 
Starting from these multi-bubble states, we could further identify coherent paths connecting to more exotic symmetric states. The states that are reversibly connected via these consecutive coherent two-state oscillations form an orbital structure. The coherent preparation and interference of states within this orbital would be relevant to quantum information processing and Ramsey-type interferometry-based precision measurements.

Moreover, as shown in Fig.~\ref{S3subleading}(b), once the nearest-neighbour Ising term in Eq.~(\ref{SM.eq.H}) is replaced by the $r^{-3}$ long-range interaction, the sub-leading overlap of $\vert S_3 \rangle$ exhibits more exotic resonant features. The long-range part of the interaction not only shifts the resonant longitudinal field $h$, but also lifts the degeneracy of the multi-bubble manifold (see the splitting at $h\sim 2.1$), thereby enhancing the coherence. Furthermore, the interplay between the transverse field $g$ and the long-range interaction induces a branch-splitting structure around $h \sim 2.1$. Such a non-analytic branch-splitting structure might be non-perturbative and thus challenging to capture using perturbative approaches (such as SWT) starting from the unperturbed system at $g = 0$. The intriguing orbital structure associated with the long-range interaction will be examined in future work.

\begin{figure}[p]
    \centering
    \includegraphics[width=1.0\linewidth]{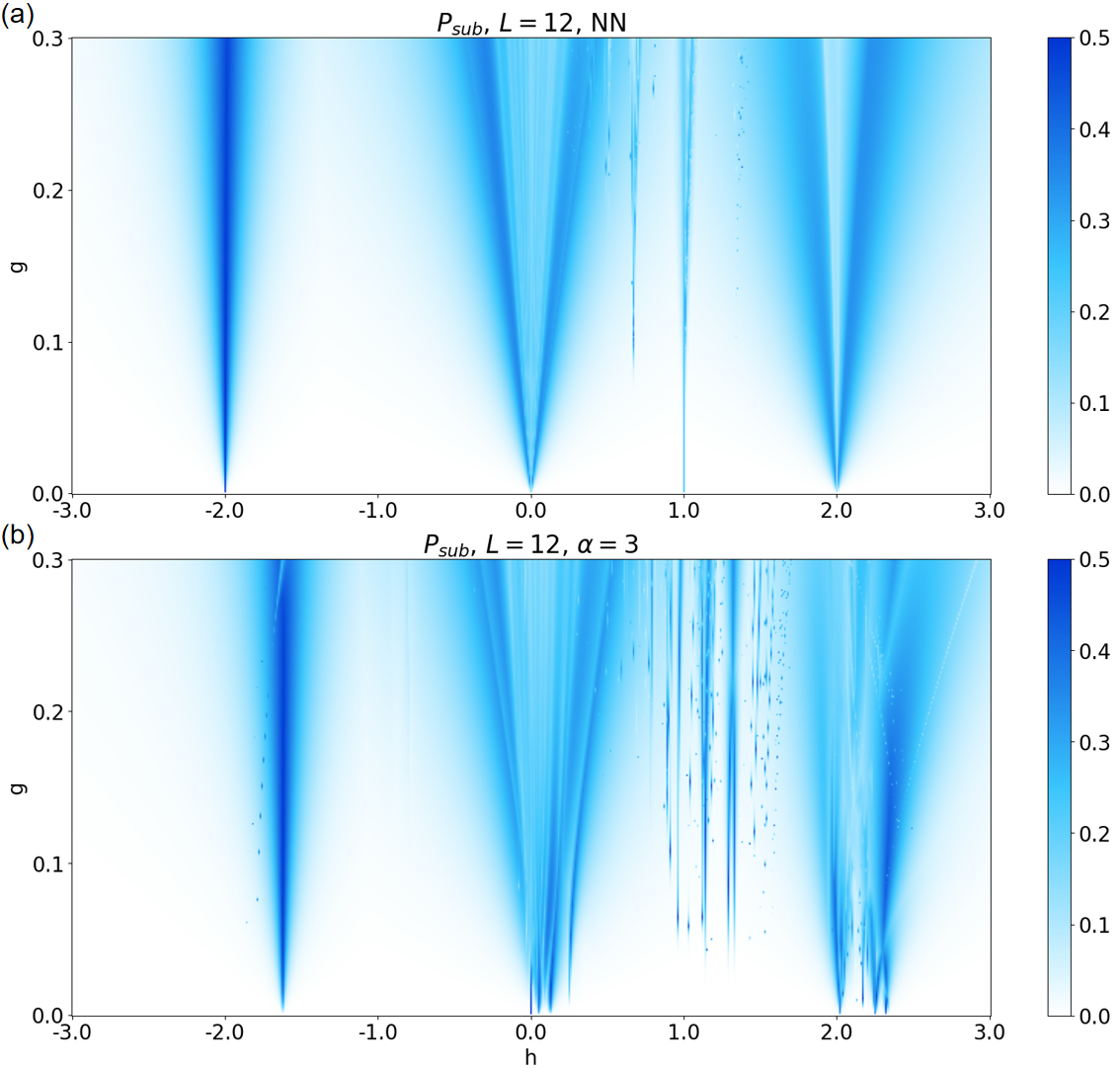}
    \caption{The same as Fig.~\ref{S3subleading} but with $L=12$.}
    \label{S3subleading_12}
\end{figure}

In Fig.~\ref{S3subleading_12}, we plot the sub-leading overlap between the symmetric state $\vert S_3 \rangle$ and the eigenstates of the TLFIM Hamiltonian for a chain of length $L=12$. The same degeneracy lifting and branch-splitting behaviors are again revealed.

\end{document}